\documentclass[final]{IEEEtran}

\usepackage{graphicx}
\usepackage{subfigure}
\usepackage{multirow}
\usepackage{array}
\newcolumntype{P}[1]{>{\centering\arraybackslash}p{#1}}
\newcolumntype{M}[1]{>{\centering\arraybackslash}m{#1}}

\usepackage{amsthm,amssymb,amsmath,color,amsfonts}%
\usepackage[update,prepend]{epstopdf}
\usepackage{multirow}
\usepackage[latin1]{inputenc}
\usepackage{tikz}
\usepackage{bbm} % for \mathbbm{1}
\usepackage{pdfpages}
\usepackage{multirow}
\usepackage{subfig}
\usepackage{comment}
\usepackage{graphicx}
\usepackage{multicol}
\usepackage{cite}
\usepackage{graphicx}
\usepackage{subcaption}

\usepackage[justification=centering]{caption}
\usepackage{textcomp}
\usepackage{psfrag}
\usepackage{arydshln}
\usepackage{url}
\usepackage{soul}
\usepackage{graphicx,color}
\usepackage[nolist]{acronym}
\usepackage{algorithm,algorithmic} %algorithm
\usepackage{mathtools,lipsum}
\usepackage{cuted}
\usepackage{mathrsfs}

\usepackage[capitalise]{cleveref}
\Crefname{equation}{Eq.\!}{Eqs.\!}
\Crefname{figure}{Fig.\!}{Figs.\!}
\Crefname{tabular}{Tab.\!}{Tabs.\!}
\Crefname{section}{Section\!}{Sections.\!}

\def\nb0{{\mathbf{0}}}
\def\nb1{{\mathbf{1}}}

\newtheorem{lemma}{Lemma}

\newtheorem{definition}{Definition}

\newtheorem{theorem}{Theorem}

\newenvironment{sequation}{
\begin{equation}\small}{\end{equation}
}

\begin{document}
%\pagenumbering{gobble}
\graphicspath{{./Figures/}}

\title{Performance Analysis of SAGIN from the \\ Relay Perspective: A Spherical Stochastic \\ Geometry Approach}

%Performance Analysis of a Hybrid RF/FSO Satellite-Aerial-Terrestrial Network from a Relay Perspective: A Stochastic Geometry Approach

\author{Ferdaous Tarhouni, Ruibo Wang, {\em Member, IEEE}, and Mohamed-Slim Alouini, {\em Fellow, IEEE}

\thanks{The authors are with the Computer, Electrical and Mathematical Science and Engineering Division (CEMSE), King Abdullah University of Science and Technology (KAUST), Thuwal 23955-6900, Makkah Province, Saudi Arabia (e-mail: ferdaous.tarhouni@kaust.edu.sa; ruibo.wang@kaust.edu.sa; slim.alouini@kaust.edu.sa). Corresponding author: Ruibo Wang.}
\vspace{-8mm}
}

\maketitle

\begin{abstract}
In recent years, the satellite-aerial-ground integrated network (SAGIN) has become essential in meeting the increasing demands for global wireless communications. In SAGIN, high-altitude platforms (HAPs) can serve as communication hubs and act as relays to enhance communication performance. In this paper, we evaluate network performance and analyze the role of HAPs in SAGIN from the relay perspective. Based on this unique perspective, we introduce three metrics to evaluate the performance, named the average access data rate, the average backhaul data rate, and the backhaul rate exceedance probability (BREP). Considering the need for dynamic topology and interference analysis, we choose spherical stochastic geometry (SSG) as a tool and derive analytical expressions for the above metrics to achieve low-complexity performance evaluation. Specifically, we provide a closed-form expression for the end-to-end performance metric BREP. Given that there is no existing literature in the SSG field studying networks from a relay perspective, we specifically investigate the impact of satellite network topology on performance in our numerical results to further highlight the advantages of the SSG framework. Additionally, we analyze the minimum HAP transmission power required to maintain both short-term and long-term data rate demands.
\end{abstract}


\vspace{-1mm}

\section{Introduction}
\subsection{Motivation}
%\IEEEPARstart{T}{he}
Recently, both academia and industry have sparked significant interest in advanced air-ground integration strategies \cite{9822386}. The satellite-aerial-ground integrated network (SAGIN) is recognized as a promising architectural paradigm for future sixth-generation (6G) wireless communications \cite{9466942}. With its advantages of extensive coverage, high throughput, and robust resilience, SAGIN is ideal for various applications, including Earth observation, intelligent transportation systems, and disaster relief missions \cite{8368236}. Direct links between ground users and satellites are often hindered by long transmission distances and severe path loss, especially in uplink communication where user power is limited. Aerial relays in SAGIN, such as high-altitude platforms (HAPs), can effectively bridge communication between ground users and satellite networks \cite{8438489}. HAPs can maintain a quasi-stationary position relative to the ground for extended durations \cite{10355104}, support a higher payload capacity for communication units \cite{11006480}, and can achieve line-of-sight connectivity with a higher probability \cite{huang2024system,10821003}. As a result, the air-ground-space integrated cross-layer communication approach can effectively enhance communication reliability and improve network resilience.

\par
With the large-scale deployment of HAPs, UAVs, and satellites, network performance evaluation has become increasingly important. Accurate evaluation results can provide operators with key recommendations for network configuration and resource allocation. Authors of \cite{tan2024outage} studied the uplink performance of a SAGIN, focusing on the joint impact of UAV altitude, position angle, and energy limitations. They derived closed-form expressions for energy and signal-to-noise ratio (SNR) outage probabilities, and transmission time gap in ground-to-air-to-space and ground-to-space communications. In \cite{wei2024energy}, the authors proposed a joint caching optimization and user selection (JCOUS) framework for UAV-enabled SAGINs, to maximize the residual energy of satellites while satisfying UAV resource constraints.
Meanwhile, authors of \cite{jia2025service} presented a dynamic service function chain (SFC) scheduling strategy for SAGINs, integrating heterogeneous computing resources across the space, air, and ground layers. By leveraging the SFC structure and exploiting node mobility, the authors introduced a deep reinforcement learning algorithm to dynamically assign network functions while minimizing long-term transmission delay and operational costs.

So far, multiple studies have shed light on the significant role of HAP particularly as a relay to enhance satellite-ground communications and analyzed the communication performance of the user-HAP-satellite architecture. In \cite{9933613}, authors demonstrated the advantages of employing a HAP as a relay to enhance communication reliability by evaluating and comparing outage probabilities. The system throughput and average frame delay were derived for a HAP-assisted relaying satellite-to-vehicle FSO network in \cite{9625366}. In \cite{bithas2024hybrid}, the authors analyzed the performance of a hybrid FSO/RF downlink communication within a SAGIN architecture using a HAP as a relay. A novel low-overhead channel selection scheme is proposed to reduce signaling complexity during link switching. The authors derived exact and asymptotic expressions for the end-to-end SNR and evaluated system performance in terms of outage probability, switching probability, and the average number of link estimations using Markov chain theory. A distributed decode-and-forward relaying strategy with multiple HAPs has been used in \cite{bithas2024distributed}. The HAPs that successfully decode the satellite's FSO signal collaboratively forward it to the ground station over RF links using a distributed space-time coding (DSTC) scheme. The authors derived both exact and asymptotic analytical expressions for the outage probability. On the other hand, authors of \cite{9314201} investigated the uplink transmission in a hybrid RF/FSO integrated terrestrial-aerial-space network, where a HAP serves as a relay node. This paper developed a space division multiple access and beamforming schemes to maximize the ergodic sum rate. 

\par
Although some of the aforementioned studies acknowledge the crucial role of HAP as a relay hub in SAGINs, they still exhibit limitations. First, a HAP typically covers a vast area with a high user density, where inter-user communication interference cannot be ignored \cite{shamsabadi2022handling}. Second, both HAPs and LEO satellites move rapidly relative to the ground \cite{azari2022evolution}. Neglecting the dynamic positional changes of satellites can lead to inaccurate performance estimations of the HAP-satellite link due to fluctuations in distance. The mainstream approach to addressing the above two challenges is numerical simulation \cite{jiang2023network,roshdi2024performance}. However, for large-scale HAP or satellite networks, the computational complexity of numerical simulations is extremely high \cite{10278941}. Although numerical simulations can model interference links, the number of links typically grows quadratically with the number of devices. Additionally, numerical simulations require dynamic modeling of device positions based on network configurations, such as simulating the movement of satellites along their orbits. Due to the dynamic changes in network topology and the randomness of the channel, at least thousands of simulation rounds are often needed to obtain relatively accurate performance evaluation results.

\subsection{Related Works}
Notably, stochastic geometry (SG) is considered one of the most suitable analytical methods for modeling large-scale dynamic networks and possesses unique interference analysis capabilities \cite{okati2020downlink}. In addition to addressing the challenges of dynamic topology and interference analysis, the delay in performance evaluation using the SG framework has been proven to be less than one-four-thousandth of that of state-of-the-art numerical simulation methods \cite{10278941}. Through mathematical derivation, the SG analysis framework expresses performance metrics as analytical functions of network parameters. With these analytical expressions, we can directly map network parameters, such as devices' altitude, to performance metrics such as coverage probability \cite{al2021optimal}, therefore significantly reducing computational delay.

\par
Several SG-based studies explore the performance of SAGINs. In \cite{10438999}, the authors proposed an analytical framework based on spherical stochastic geometry (SSG) to investigate the uplink path connectivity in a SAGIN, where aerial vehicles were employed as relay nodes. The outage performance of a cooperative satellite-aerial-terrestrial network was investigated in \cite{9841465}, where a cluster of aerial platforms aims to transmit data to a terrestrial terminal through two types of relaying links: one via a satellite and the other via aerial nodes. Authors of \cite{8951059} investigated the outage performance as well, for a hybrid satellite-UAV terrestrial network, where a UAV acts as a relay to forward signals to ground users. Moreover, the authors determined the optimal UAV placement to maximize the sum rate. In \cite{9130899}, the authors analyzed the downlink performance of a cooperative satellite-aerial-terrestrial network, considering aerial platforms employed as relays using the decode-and-forward scheme. 

\par
On the other hand, the secrecy outage performance of SAGINs has been explored \cite{9919807,10465173,aerospace11040306} under the SG framework. In particular, authors of \cite{9919807} and \cite{aerospace11040306} derived end-to-end secrecy outage probability for an uplink transmission within a hybrid RF/FSO SAGIN. Moreover, some studies have specifically examined dual-hop relay communication. In \cite{10580980}, the authors investigated uplink communication in a terrestrial-satellite network where a LEO satellite serves as a relay, forwarding signals from IoT devices to a geosynchronous equatorial orbit (GEO) satellite. In addition, the authors in \cite{9586571} investigated dual-hop communication within a hybrid RF/FSO cooperative satellite-UAV network. In their framework, aerial cluster heads acted as DF relays connecting the satellite and UAVs. Analytical and asymptotic expressions were derived for the coverage probability under various scenarios: interference-free, interference-dominated, and interference-plus-noise.

\par
Although the performance analysis framework for SAGINs based on SG has been developed to some extent, it is still worth further refinement. Firstly,  previous works predominantly consider UAVs as aerial relay platforms, whereas the potential benefits of employing HAPs have not been adequately investigated. Secondly, a significant part of the prior studies assumes a simplified scenario involving only one satellite. Consequently, the fixed relative position between the user and the satellite makes the performance evaluation no longer universally meaningful and does not reflect real-world operational conditions. Last but most importantly, there is currently no literature in the field of SSG that analyzes network performance from a relay perspective. This means that many key issues may remain unanswered, such as how to control the power of HAPs to meet the data rate requirements of users. 

\par
Finally, we explain the challenges brought to this paper by overcoming the above limitations. Introducing HAPs as relays and modeling the entire satellite constellation will increase the difficulty of channel and topology analysis. Moreover, evaluating network performance from a relay perspective often involves end-to-end analysis of performance metrics. For instance, in uplink communication, the performance distribution of the HAP-satellite link is based on the conditional distribution of the user-HAP link performance. As a result, the analytical difficulty is much higher than in the case where two-hop links are independent.

\subsection{Contributions}
The key contributions of this paper can be summarized as follows.
\begin{itemize}

\item \textit{Relay Perspective Analysis:}
To the best of our knowledge, this is the first work in the context of spherical SG that investigates network performance from the relay perspective. Specifically, we focus on the design and performance analysis of the HAP as a relay. To do so, we investigate how to configure its transmit power and location to ensure that the backhaul link satisfies users' uplink rate requirements. This contrasts with the conventional user perspective, adopted in literature, which emphasizes metrics such as coverage probability or user data rates.

\item \textit{Metric Design:}
We introduce three performance metrics for the relay perspective: average access data rate (AADR), average backhaul data rate (ABDR), and backhaul rate exceedance probability (BREP), capturing both short-term and long-term network behavior.

\item \textit{Closed-Form Expression:}
We derive analytical expressions for all proposed metrics. Notably, the end-to-end performance metric (BREP) is obtained in closed form, offering significant computational advantages, an uncommon result given the inherent complexity of spherical SG models.

\item \textit{Topology Impact and Power Optimization:}
We conduct a comprehensive analysis of how satellite network topology affects the proposed metrics. Additionally, we optimize the HAP transmit power to improve both short and long-term performance, demonstrating the practical utility of our framework for system design.
\end{itemize}

The remainder of this paper is structured as follows. The system model of the considered SAGIN is described in Sec.~\ref{2}. In Sec.~\ref{3}, the topological analysis and the analytical expressions for the performance metrics are explored. The numerical results using Monte Carlo simulations are presented and discussed in Sec.~\ref{4}. Some future research directions are outlined in Sec.~\ref{research}. Finally, the concluding remarks are given in Sec.~\ref{5}.

\section{System Model}
\label{2}
This section describes the network's spatial distribution model and discusses the channel models used for user-HAP and HAP-satellite communication links.

\begin{figure}[ht]
\centering
\includegraphics[width=\linewidth]{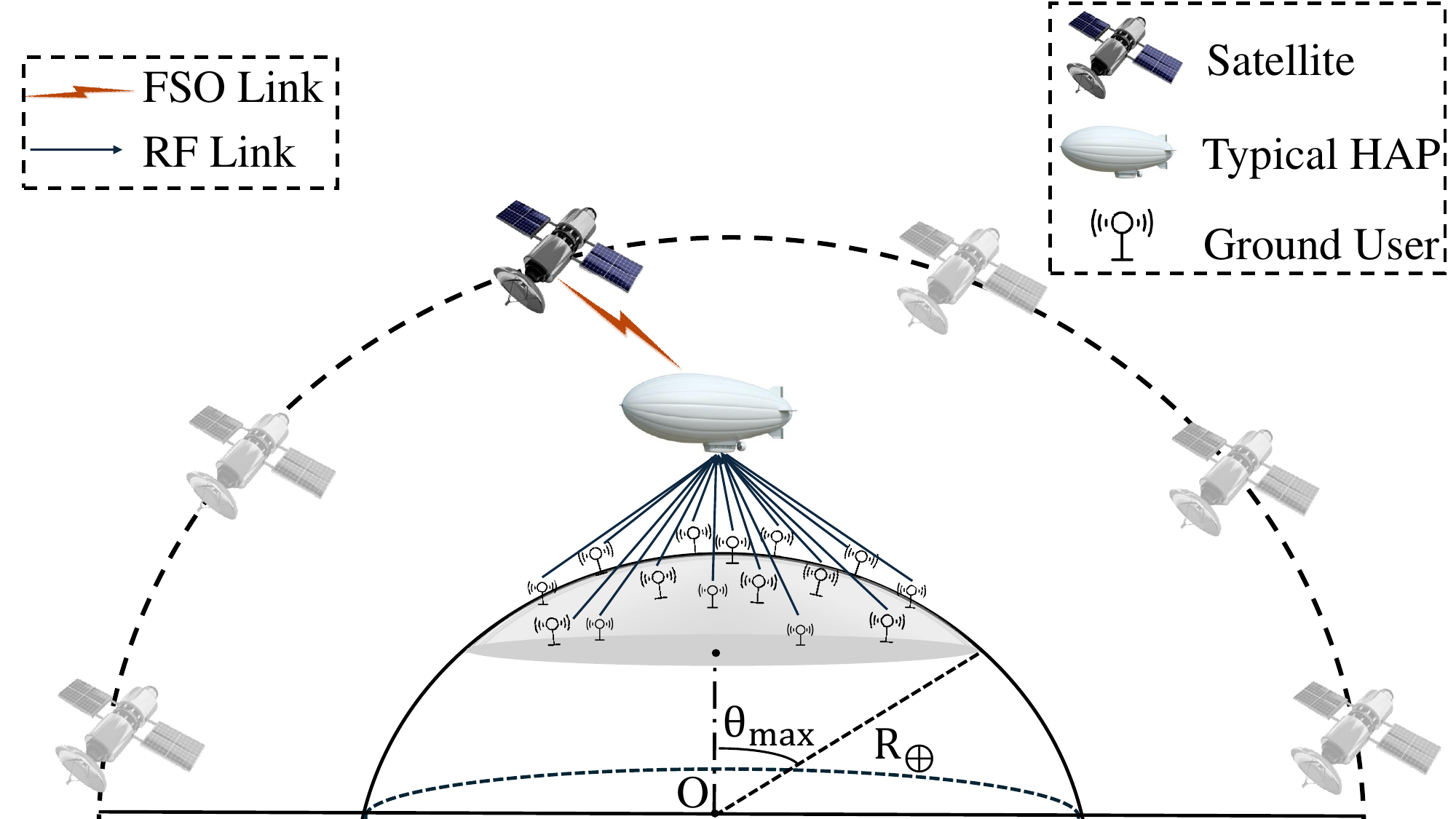}
\caption{System model of the considered SAGIN.}
\label{system}
\end{figure}

\subsection{Spatial Distribution Model}
We consider a SAGIN as given in Fig.~\ref{system}, where users transmit signals to LEO satellites via HAPs as relays \cite{9314201}, \cite{huaicong2022ergodic}. We take the Earth center as the origin and denote the Earth radius as $R_\oplus$. The positions of ground users follow a spherical homogeneous Poisson point process (PPP) $\Phi_u$ with density $\Lambda_u$\cite{10082988}. $N_s$ LEO satellites are distributed according to a spherical homogeneous binomial point process (BPP), denoted as $\Phi_s$ \cite{wang2023reliability}, \cite{10930516}. The radius of the sphere on which the satellites are distributed is denoted as $R_s$. HAPs are located on a sphere with a radius $R_H=R_\oplus+H$, where $H$ is the altitude of HAPs. We evaluate the network performance from the perspective of a typical HAP. Due to the rotational invariance of the homogeneous point processes \cite{240305}, the coordinates of this typical HAP are, without loss of generality, denoted as $(R_H,0,0)$. The typical HAP decodes the signals from ground users and forwards them to the satellite with the strongest average received power.

\subsection{User-HAP Channel Model}
Consider that users communicate with the typical HAP through the RF band, where the channel experiences both path loss and small-scale fading. The received power from the $i$-th user at the HAP is given as \cite{9919807}:
\begin{equation}
\label{powerHi}
    \rho_{H,i}^r = P_u G_u G_H^r d_i^{-\alpha} \left( \frac{\lambda_{\mathrm{RF}} h_{\mathrm{RF}}}{4 \pi} \right)^2,
\end{equation}
where $P_u$ is the transmission power of the user, $G_u$ and $G_H^r$ are the user's transmission antenna gain and the HAP's reception antenna gain. $d_i$ denotes the distance between the $i$-th user and the HAP. $d_i \leq d_{\max}$ should be satisfied for $\forall i$ to ensure that users are within the HAP's service range and that the user-HAP links are not obstructed by the Earth. $\alpha$ is the path loss exponent, and $\lambda_{RF}$ is the radio wavelength.  

\par
$h_{\mathrm{RF}}$ denotes the channel coefficient and is assumed to follow the shadowed Rician fading, which is considered one of the most accurate models for terrestrial-aerial links \cite{9314201}. The probability density function (PDF) of $h_{\mathrm{RF}}^2$ can be approximately expressed as \cite{9511625}:
\begin{equation}
    f_{h_{\mathrm{RF}}^2}(h) \approx \frac{1}{m_2^{m_1} \Gamma\left(m_1\right)} h^{m_1-1} \exp \left(-\frac{h}{m_2}\right),
\end{equation}
where $\Gamma\left( \cdot \right)$ is the Gamma function. $m_1$ is the shape parameter and $m_2$ is the scale parameter. Finally, when the HAP receives a signal from a particular user, the signals from other users are considered interference.

\subsection{HAP-Satellite Channel Model}
Considering the high data rate requirements of the backhaul link, the HAP-satellite link employs FSO transmission \cite{9586571}. In contrast, users may be more inclined to use the RF band for cost considerations. Therefore, the received power of the satellite is modeled as \cite{9919807}:
\begin{equation}
\label{powerFSO}
    \rho_s = \upsilon^2 P_H^t G_H^t G_s \left(\frac{ \lambda_{\mathrm{FSO}} h_{\mathrm{FSO}}}{4\pi d_s} \right)^2,
\end{equation}
where $\upsilon$ is the optical-to-electrical conversion coefficient, and $P_H^t$ is the transmission power of the HAP. $G_H^t$ and $G_s$ are respectively the HAP transmission antenna gain and satellite reception antenna gain. $\lambda_{\mathrm{FSO}}$ is the optical wavelength. $d_s$ is the distance between the typical HAP and its associated satellite. 

\par
In (\ref{powerFSO}), $h_{\mathrm{FSO}}$ is the channel fading coefficient of the FSO link. 
For the HAP-satellite link, the impact of atmospheric turbulence is minimal since the laser beam propagates through a non-atmospheric path above $20$~km \cite{9625366}. As a result, misalignment fading is considered the primary impairment for this link \cite{10930516}. Given the deviation angle of the beam as $\theta_d$, the conditional PDF of $h_{\mathrm{FSO}}$ can be given by \cite{9933613}: 
\begin{equation}
f_{h_{\mathrm{FSO}}|\theta_d}\left(h\right) =
\frac{\eta_s^2}{A_0^{\eta_s^2}} h^{\eta_s^2 -1} \cos \theta_d , \ 0 \leq h \leq A_0,
\end{equation}
where $\eta_s$ and $A_0$ are parameters of the pointing error. The deviation angle $\theta_d$ is modeled by Rayleigh distribution with variance $\sigma_0^2$ and its PDF is given by \cite{9933613}:
\begin{equation} f_{\theta_d}\left(\theta\right)=\frac{\theta}{\sigma_0^2} \exp\left(-\frac{\theta^2}{2\sigma_0^2} \right), \ \theta_d \geq 0.
\end{equation}

Since the FSO link is point-to-point communication, we neglect the interference caused by other HAPs associated with the same satellite.

\section{Performance Analysis}
\label{3}
This section starts with the topological analysis. Then, we present preliminaries about interference and signal-to-interference noise ratio (SINR). Finally, we derive analytical expressions for performance metrics, including AADR, ABDR, and BREP.

\subsection{Topological Analysis}
This subsection presents topological-level results, including distance distributions of RF and FSO links. For RF links, we focus on the link from every user to the HAP.

\begin{lemma}\label{lemma1}
The PDF of the distance between the $i$-th user and the typical HAP is given by:
\begin{equation}
f_{d_i}(d)= \frac{d}{R_\oplus R_H(1-\cos\theta_{\max})}, \quad  H \leq d \leq d_{\max},
\end{equation}
where $\theta_{\max}$ is defined as
\begin{equation}
\theta_{\max} = \arccos \left( \frac{R_\oplus^2 + R_H^2 - d_{\max}^2}{2 R_\oplus R_H} \right).
\end{equation}
\end{lemma}
\begin{IEEEproof}
    See Appendix~\ref{app:lemma1}.
\end{IEEEproof}

Noting that due to the homogeneity and independence of user distribution, $f_{d_i}(d)$ is not a function of $i$. Based on the strongest average received power association, the nearest satellite is associated. Furthermore, the analytical expression of FSO link's distance distribution is given in the following lemma.

\begin{lemma}\label{lemma2}
The PDF of the distance between the typical HAP and its nearest satellite is given by:
\begin{equation}
f_{d_s}(d) = \frac{N_s d}{2 R_H R_s} \left(\frac{(R_H+R_s)^2-d^2}{4R _H R_s}\right)^{N_s-1},
\end{equation}
where $R_s - R_H \leq d \leq \sqrt{R_s^2 - R_\oplus^2} + \sqrt{R_H^2 - R_\oplus^2}$.
\end{lemma}
\begin{IEEEproof}
    See Appendix~\ref{app:lemma2}.
\end{IEEEproof}

\par
In the above expression, we have considered the scenario where communication is interrupted due to the HAP-satellite link being blocked by the Earth.

\subsection{Interference and SINR}
In this subsection, we first denote the total received power from all the users as $\mathcal{I}$:
\begin{equation} \label{sump}
\mathcal{I}=\sum_{i\in \Phi_u} \rho_{H,i}^r.
\end{equation}
In addition, the cumulative interference power of the $i$-th user from the other users at the serving HAP is denoted as $\mathcal{I}_i$:
\begin{equation}
\mathcal{I}_i=\sum_{j\in \Phi_u\setminus i} \rho_{H,j}^r = \mathcal{I} - \rho_{H,i}^r.
\end{equation}

\par
Then, we derive the Laplace transform of $\mathcal{I}$, which will serve as an important lemma for the analysis of performance metrics.

\begin{lemma} \label{lemma3}
The Laplace transform of the total received power at the typical HAP is
\begin{equation}
\begin{split}
& \mathcal{L}_{\mathcal{I}}(s)= \exp \Bigg(-2\pi\Lambda_u R_\oplus^2 \\
& \times \int_0^{\theta_{\max}} 
\Bigg[1 - \Bigg(1+sm_2 P_{u}G_{u}G_H^r \left( \frac{\lambda_{\mathrm{RF}}}{4\pi}\right)^2 \\
& \times (R_\oplus^2+R_H^2-2R_\oplus R_H \cos\theta)^{-\frac{\alpha}{2}} \Bigg)^{-m_1} \Bigg] 
\sin\theta \, \mathrm{d} \theta \Bigg).
\end{split}
\end{equation}
\end{lemma}
\begin{IEEEproof}
    See Appendix~\ref{app:lemma3}.
\end{IEEEproof}

\par
In the second part of this subsection, we focus on the SINR of the RF link and the signal-to-noise-ratio (SNR) of the FSO link. According to the definition, the SINR of the $i$-th user at the typical HAP can be written as: 
\begin{equation}\label{SINRRF}
\gamma_{H,i}=\frac{\rho_{H,i}^r}{\sigma_H^2 + \mathcal{I}_i},
\end{equation}
where $\sigma_H^2$ is the noise power at the HAP. Recall that the interference is ignored, as explained in the last section. Similarly, the SNR of the FSO link at the satellite is defined as $\gamma_s = \rho_s / \sigma_s^2$, where $\sigma_s^2$ is the noise power at the satellite.

\par
Finally, we define a random variable related to $\gamma_{H,i}$ as
\begin{equation}
\mathcal{Z} = \frac{\rho_{H,i}^r}{\sigma_H^2 + \mathcal{I}},
\end{equation}
and conclude this subsection by deriving the complementary cumulative density function (CCDF) of $\mathcal{Z}$.

\begin{lemma} \label{lemma4}
The CCDF of $\mathcal{Z}$ is derived as follows:
\begin{equation}
\begin{split}
& \overline{F}_{\mathcal{Z}}(z) = \sum_{k=1}^{m_1} \binom{m_1}{k} (-1)^{k+1} \\
& \times \exp(-k\beta\mu(d_i)\sigma_H^2) \mathcal{L}_{\mathcal{I}} \bigg(k \beta \mu(d_i)\bigg),
\end{split}
\end{equation}
where $0 \leq z \leq 1$. $\beta$ is defined as
\begin{equation}
    \beta=\left(m_1!\right)^{-\frac{1}{m_1}},
\end{equation}
and $\mu(d_i)$ is expressed as
\begin{equation}
    \mu(d_i)=\frac{z}{m_2}\left(\frac{4\pi}{\lambda_\mathrm{RF}}\right)^2\frac{1}{P_u G_u G_{H}^r} d_i^{\alpha}.
\end{equation}
\end{lemma}
\begin{IEEEproof}
    See Appendix~\ref{app:lemma4}.
\end{IEEEproof}

\par
Based on the four given lemmas, the network performance metrics can be derived.

\subsection{Average Data Rates}
Now, we provide the definition and analytical expressions for the average access and the average backhaul data rates.

\begin{definition}
AADR is defined as the average overall data rates of all user-HAP links. Mathematically, AADR can be expressed as: 
\begin{equation}
\mathcal{R}_{\mathrm{RF}} = \mathbb{E} \left[ \sum_{i\in \Phi_u} B_{\mathrm{RF}}\log_2\left(1+\gamma_{H,i}\right) \right],
\end{equation}
where $B_{\mathrm{RF}}$ denotes the RF bandwidth, and $\gamma_{H,i}$ is the instantaneous SINR defined in (\ref{SINRRF}).
\end{definition}

Then, the analytical expression of AADR is given in the following theorem.

\begin{theorem}\label{theorem1}
The AADR is given by:
\begin{equation}
\mathcal{R}_{\mathrm{RF}} = \frac{ 2\pi \Lambda_u R_\oplus^{2} B_{\mathrm{RF}} }{\ln2} \int_{0}^{\theta_{\max}} \sin\theta \int_{0}^{1} \overline{F}_{\mathcal{Z}}(z) \, \mathrm{d}z \, \mathrm{d} \theta, 
\end{equation}
where $\overline{F}_{\mathcal{Z}}(z)$ is defined in Lemma~\ref{lemma4}.
\end{theorem}
\begin{IEEEproof}
    See Appendix~\ref{app:theorem1}.
\end{IEEEproof}

Next, ABDR can be defined and derived similarly.

\begin{definition}
ABDR is defined as the average data rate of the HAP-satellite link. Mathematically, ABDR can be expressed as:
\begin{equation}
\mathcal{R}_{\mathrm{FSO}}= \mathbb{E} \left[ B_{\mathrm{FSO}}\log_2\left(1+\gamma_s\right) \right],
\end{equation}
where $B_{\mathrm{FSO}}$ denotes the FSO bandwidth and $\gamma_s$ is the SNR at the associated satellite.
\end{definition}

Before deriving ABDR, we need the PDF of the FSO link channel fading coefficient as a lemma.

\begin{lemma} \label{lemma5}
The PDF of $h_{\mathrm{FSO}}$ is given as follows:
\begin{sequation}
f_{h_{\mathrm{FSO}}} \left(h\right) = \frac{\eta_s^2}{A_0^{\eta_s^2}} h^{\eta_s^2-1} 
\exp\left(-\frac{\sigma_0^2}{2}\right) {}_1F_1\left( \frac{-1}{2}, \frac{1}{2}, \frac{\sigma_0^2}{2} \right),
\end{sequation}
where $0 \leq h \leq A_0$, and ${}_1F_1( \cdot, \cdot, \cdot )$ is the confluent Hypergeometric function.
\end{lemma}
\begin{IEEEproof}
    See Appendix~\ref{app:lemma5}.
\end{IEEEproof}

Based on $f_{h_{\mathrm{FSO}}} \left(h\right)$, the analytical expression of ABDR can be derived in the following theorem.

\begin{theorem}\label{theorem2}
The ABDR is given by
\begin{equation}
\begin{split}
& \mathcal{R}_{\mathrm{FSO}} = \frac{B_{\mathrm{FSO}}}{\ln2}\int_{R_s-R_H}^{d_{\mathrm{H2s}}^{\max}} \int_{0}^{A_0} f_{h_{\mathrm{FSO}}} (h) f_{d_s}(t) \\
& \times \ln \left( 1 + P_H^t G_H^t G_s \left( \frac{\upsilon  \lambda_{\mathrm{FSO}} h}{4 \pi \sigma_s t} \right)^2 \right) \mathrm{d}h \mathrm{d}t,
\end{split}
\end{equation}
where $d_{\mathrm{H2s}}^{\max}$ is defined as
\begin{equation}
d_{\mathrm{H2s}}^{\max} = \sqrt{R_H^2 - R_{\oplus}^2} + \sqrt{R_s^2 - R_{\oplus}^2}.
\end{equation}
\end{theorem}
\begin{IEEEproof}
    See Appendix~\ref{app:theorem2}.
\end{IEEEproof}

\par
So far, AADR and ABDR remain separate discussions of RF or FSO links. Next, we introduce BREP as a metric to measure end-to-end communication performance.

\begin{table*}[h!]
\centering
\renewcommand{\arraystretch}{1.3} % Increase row height by 50%
\caption{Simulation Parameters \cite{240305}, \cite{10082861}, \cite{10082988}, \cite{10227357}.}
\scalebox{1}{%
\begin{tabular}{|c|c|c||c|c|c|}
\hline
\textbf{Notation} & \textbf{Meaning} & \textbf{Default Value} &\textbf{Notation} & \textbf{Meaning} & \textbf{Default Value} \\ \hline \hline
$R_{\oplus}$        & Radius of the Earth   & $6371$~km & $P_u$     & Transmit power of user  &  $20$~W  \\ \hline
$R_s$   & Radius of satellites & $6871$~km &  $G_u$  & User gain  &    $8$~dBi  \\ \hline 
$H$  & Height of the HAP & 20 km & $G_{H}^r$     & HAP receiver antenna gain &    $32$~dBi  \\\hline
$r_H$ & HAP Coverage radius & $80$~km &  $G_{H}^t$     & HAP transmitter antenna gain &   $52$~dBi \\ \hline
$\Lambda_u$     & Density of the PPP   &   $10^{-5}/$m$^2$ & $G_s$     & Satellite antenna gain    &    $42$~dBi \\ \hline 
$N_s$     & Number of satellites &  $300$  &   $\lambda_{\mathrm{FSO}}$     & FSO wavelength    &  $1550$~nm     \\ \hline
$\eta_s, A_0$     & Parameters of the pointing error          & $1.00526$, $0.01979$ & $f_{\mathrm{RF}}$     & RF carrier frequency &    $2$~GHz \\ \hline
$\sigma_0$      & Variance of Rayleigh distribution   & 15 mrad  & $\sigma_s^2$     & Noise power of the satellite &  $1.5\times10^{-12}$~W \\\hline
$\nu$    & Optical-to-electrical conversion coefficient &  $0.5$  &  $\sigma_H^2$     & Noise power of the HAP &     $2\times10^{-14}$ W \\ \hline
$\alpha$     & Path loss component    &   $2$   & $B_{\mathrm{RF}}$ & RF link bandwidth   &  $1$~GHz \\ \hline
$m_1, m_2$     & Shadowed Rician parameters  & $2$, $0.5$   &  $B_{\mathrm{FSO}}$     & FSO link bandwidth   &  $100$~GHz   \\ \hline 
\end{tabular}%
}
\label{simul}
\end{table*}

\subsection{Backhaul Rate Exceedance Probability}
In this subsection, we first introduce the definition of the proposed end-to-end metric. Subsequently, we provide its analytical derivation.
\begin{definition}
BREP is the probability that the instantaneous data rate of the HAP-satellite backhaul link exceeds the sum of the instantaneous data rates of the user-HAP access links. Mathematically, BREP can be expressed as:
\begin{equation}
\begin{split}
\mathcal{P}_{\mathrm{BREP}} = \mathbf{Pr} \bigg\{ & \sum_{i\in \Phi_u} B_{\mathrm{RF}}\log_2\left(1+\gamma_{H,i}\right) \\
& < B_{\mathrm{FSO}}\log_2\left(1+\gamma_s\right) \bigg\}.
\end{split}
\end{equation}
\end{definition}

The following theorem gives the analytical expression of BREP.

\begin{theorem} \label{theorem3}
BREP can be expressed as:
\begin{equation}
\begin{split}
\mathcal{P}_{\mathrm{BREP}} & = 1 - \frac{1}{A_0^{\eta_s^2}}\exp\left(-\frac{\sigma_0^2}{2}\right) \, {}_1F_1\left(-\frac{1}{2}, \frac{1}{2}, \frac{\sigma_0^2}{2}\right) \\
& \times \varepsilon^{\eta_s^2}\mathbb{E}_{d_s}(d_s^{\eta_s^2}) \sum_{n=0}^{\infty} (-1)^n \frac{\Gamma(\frac{\eta_s^2}{2}+1)}{n! \, \Gamma(\frac{\eta_s^2}{2}-n+1)} \\
&\times \exp \Bigg(-2\pi \Lambda_u R_{\oplus}^2 \int_{0}^{\theta_{\max}} \bigg( 1 \\
& - \mathbb{E} \left[ \exp \left( \frac{B_\mathrm{RF}}{B_\mathrm{FSO}}(\frac{\eta_s^2}{2}-n)\mathcal{Z} \right) \right] \bigg)\sin\theta \mathrm{d}\theta \Bigg),
\end{split}
\end{equation}
where $\varepsilon$ is defined as
\begin{equation}
    \varepsilon=\frac{4 \pi \sigma_s}{\lambda_\mathrm{FSO} v \sqrt{P_H^t G_H^t G_s}}.
\end{equation}
\end{theorem}
\begin{IEEEproof}
    See Appendix~\ref{app:theorem3}.
\end{IEEEproof}

\par
Finally, the differences between AADR, ABDR, and BREP are explained. 
\begin{itemize}
\item We need to design the parameters of the HAP to ensure that ABDR is greater than AADR and BREP exceeds a certain threshold.
\item Since HAP can temporarily store data and forward it when the FSO link performance improves. Therefore, ensuring that ABDR is greater than AADR guarantees the long-term operation of the system.
\item BREP corresponds to scenarios with higher real-time communication requirements. It requires maintaining real-time forwarding in the vast majority of cases, making its requirements stricter than the average data rate.
\end{itemize}

\section{Numerical Results}
\label{4}
In this section, we present the numerical results for the derived performance metrics. The matching between the results obtained from the $10^4$ iterations of Monte Carlo simulations (marks) performed using MATLAB and the derived analytical expressions (lines) confirms the accuracy of the analytical results. Unless otherwise specified, the parameters used in these simulations will be set to their default values, illustrated in Table~\ref{simul}.

In each round, we first generate random spatial locations of ground users and LEO satellites using PPP and BPP, respectively. Next, we generate the stochastic channel links between HAP-users and HAP-satellites, and calculate the metrics of these links. Within each round, the positions of users and satellites are fixed, and then regenerated in the next round to reflect the stochastic motion and spatial dynamics of the system. The results are then averaged over $10^4$ rounds to obtain the average data rates and the end-to-end probability.

\subsection{Average Data Rate Analysis}
\label{4A}
This subsection provides insights into how the number and altitude of satellites affect the average data rates.

\begin{figure}[ht]
\centering
\includegraphics[width=0.8\linewidth]{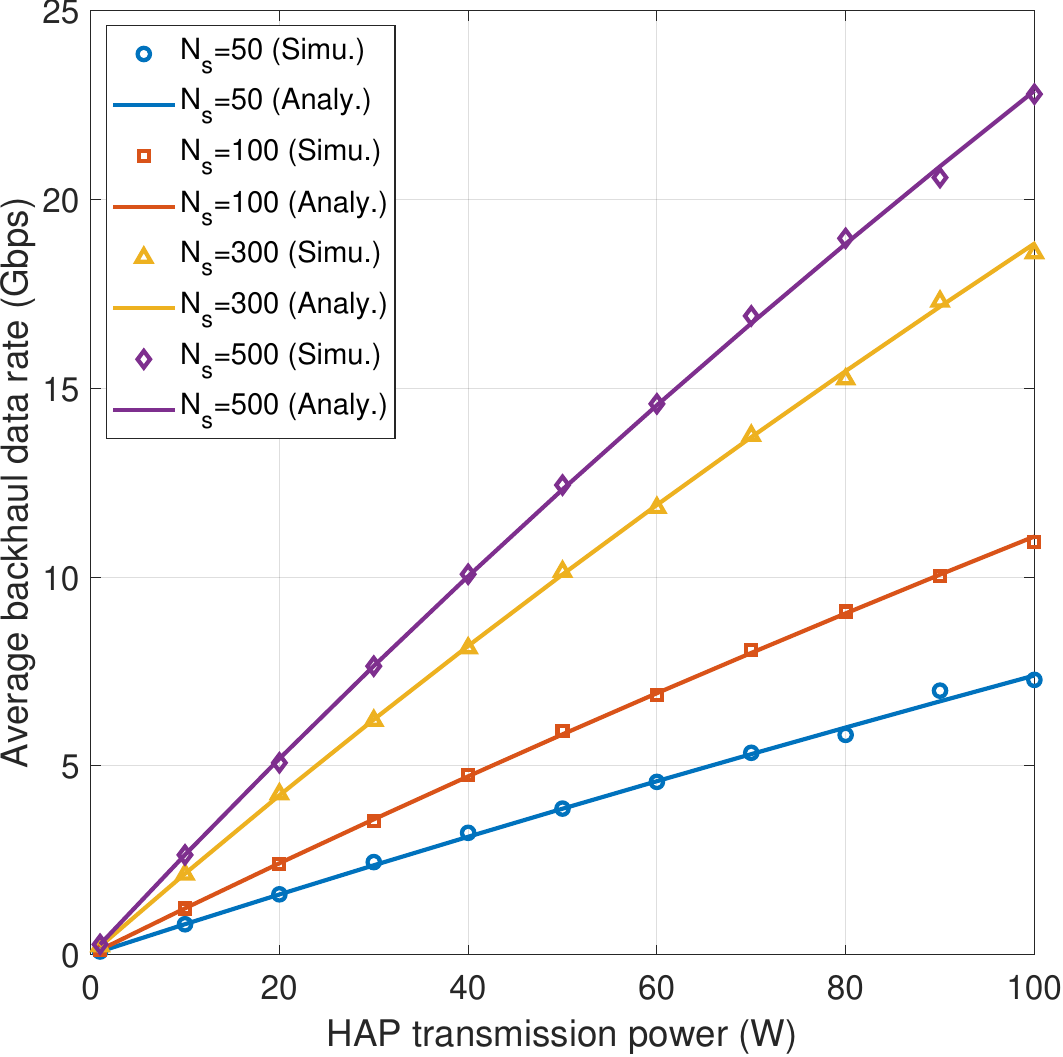}
\caption{ABDR vs $P_H^t$ for different $N_s$ values.}
\label{linedrNs}
\end{figure}

\begin{figure}[ht]
\centering
\includegraphics[width=0.8\linewidth]{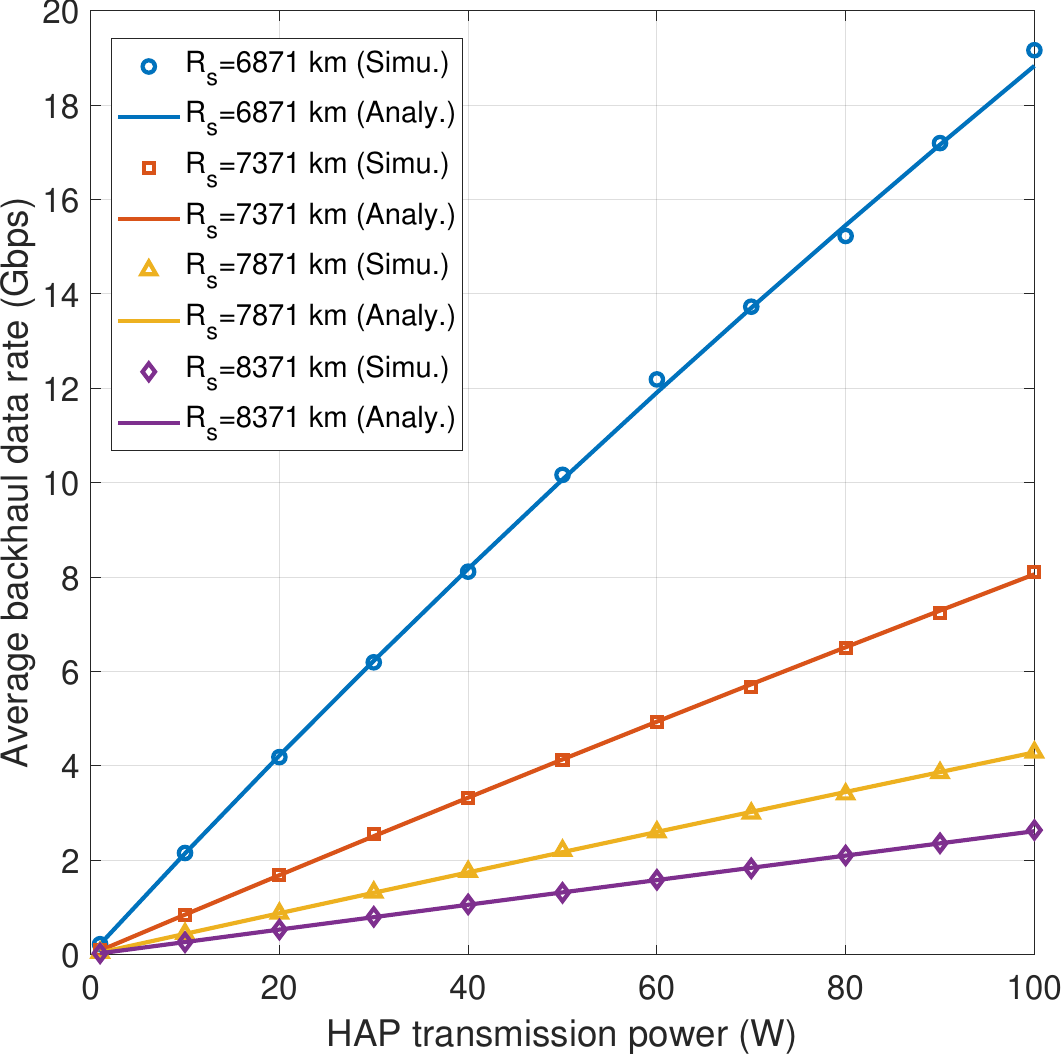}
\caption{ABDR vs $P_H^t$ for different $R_s$ values.}
\label{linedrRs}
\end{figure}

\par
In Figs.~\ref{linedrNs} and \ref{linedrRs}, the ABDR is presented for different $R_s$ and $N_s$, respectively. The transmission power of HAP and ABDR follows a simple linear relationship. This is because the HAP-satellite communication distance is long and $\gamma_s \ll 1$, thus satisfying $\ln(1+\gamma_s) \approx \gamma_s$. Additionally, the smaller the satellite radius and the greater the number of satellites, the shorter the distance from the HAP to the associated satellite, and the steeper the slope of the lines in Figs.~\ref{linedrNs} and \ref{linedrRs}.

\begin{figure}[ht]
\centering
\begin{minipage}{0.9\linewidth}
  \centering
  \includegraphics[width=\linewidth]{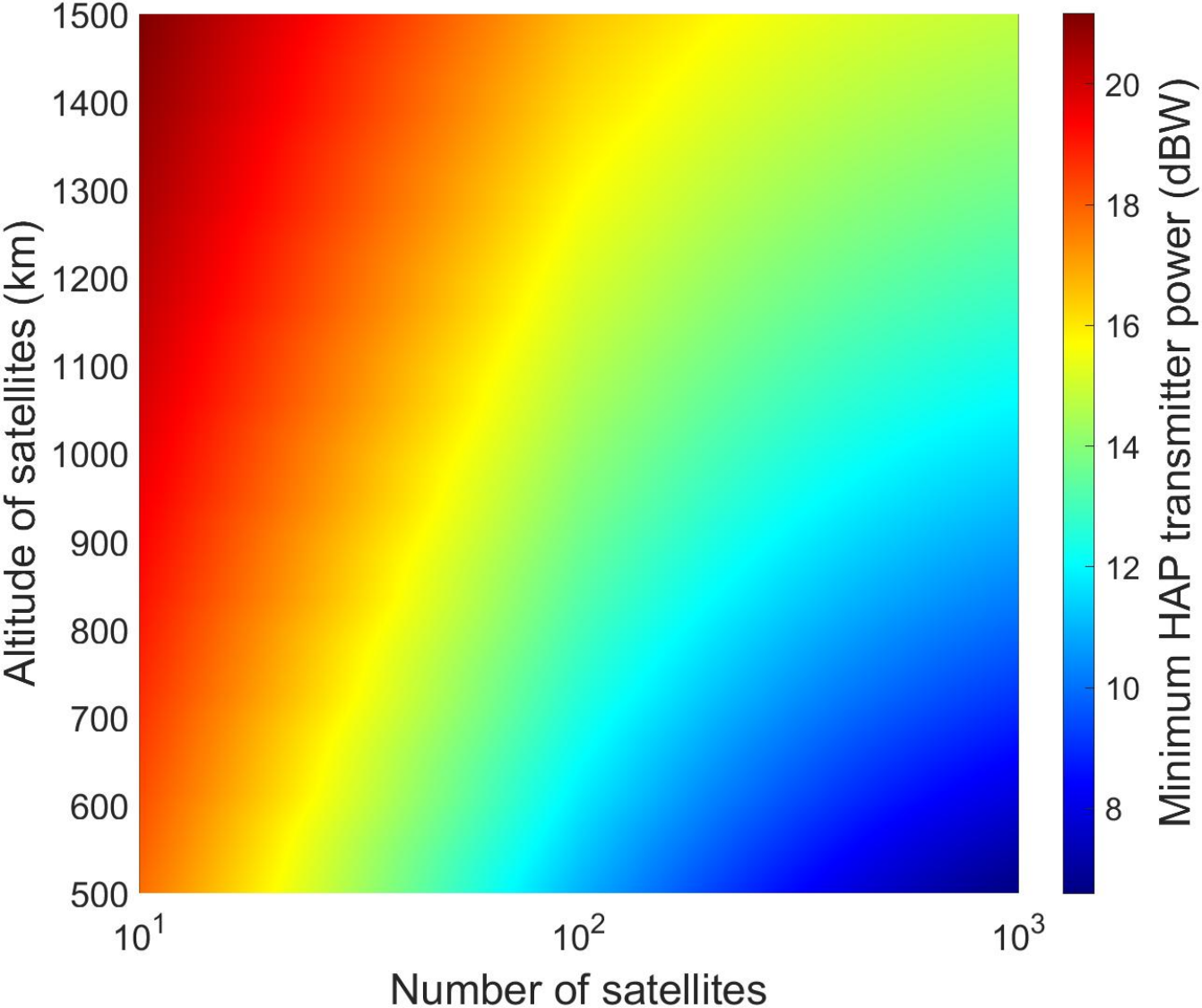}
  \captionof{subfigure}{}
  \label{heatequals}
\end{minipage}\hfill
\begin{minipage}{0.9\linewidth}
  \centering
  \includegraphics[width=\linewidth]{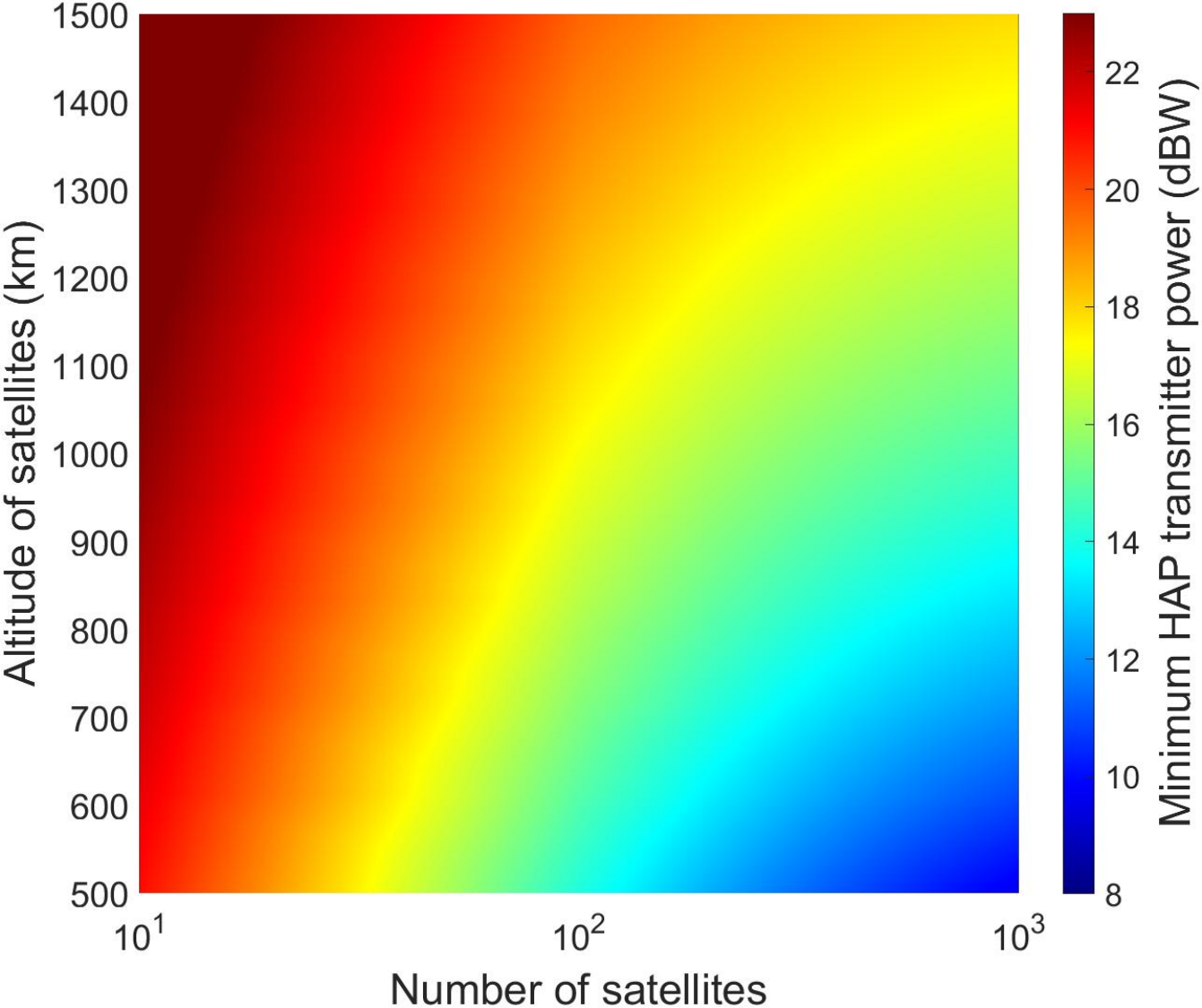}
  \captionof{subfigure}{}
  \label{heatdouble}
\end{minipage}
\caption{Minimum HAP transmission power for different constellation configurations to achieve: (a) $\mathcal{R}_{\mathrm{FSO}}=\mathcal{R}_{\mathrm{RF}}$, (b) $\mathcal{R}_{\mathrm{FSO}}=2\mathcal{R}_{\mathrm{RF}}$.}
\label{heatmaps}
\end{figure}

\par
Fig.~\ref{heatmaps} investigates the minimum $P_H^t$ required to ensure real-time communication requirements for different satellite configurations. To intuitively observe the impact of satellite height on HAP power, we define the satellite altitude as $R_s - R_{\oplus}$. Overall, reducing the satellite altitude and increasing the number of satellites can shorten the distance between HAP and its nearest satellite, therefore effectively reducing the power required by the HAP. To achieve $\mathcal{R}_{\mathrm{FSO}}=2\mathcal{R}_{\mathrm{RF}}$, the HAP obviously requires more energy than $\mathcal{R}_{\mathrm{FSO}}=\mathcal{R}_{\mathrm{RF}}$. However, the overall trend of the two heatmaps is similar.

% \begin{figure}[ht]
% \centering
% \includegraphics[width=0.9\linewidth]{HeatmapCovUser.pdf}
% \caption{Minimum HAP transmission power to achieve $\mathcal{R}_{\mathrm{FSO}}=\mathcal{R}_{\mathrm{RF}}$ under high user densities.}
% \label{heatuser}
% \end{figure}

% Fig.~\ref{heatuser} represents the minimum $P_H^t$ required to achieve equal average data rates under high user densities. It is worth mentioning that even a density of more than 100 user/km$^2$ is very high and unrealistic, but we are using this to illustrate the trend. It can be observed that higher user densities and larger HAP coverage ranges require less transmission power from the HAP to maintain balanced data rates. This happens since a higher user density leads to increased interference among users communicating with the HAP, while an expanded coverage radius increases the average distance between users and the HAP, which results in a larger path loss. Both scenarios negatively impact the SINR and, consequently, degrade the AADR. Thus, a smaller transmission power from the HAP is required.

\subsection{BREP Analysis}
\label{4B}
In this subsection, the BREP is investigated for different satellite distributions. Furthermore, we analyze the optimal HAP transmission power required to achieve specific BREPs across various satellite constellation configurations.

\begin{figure}[ht]
\centering
\includegraphics[width=0.8\linewidth]{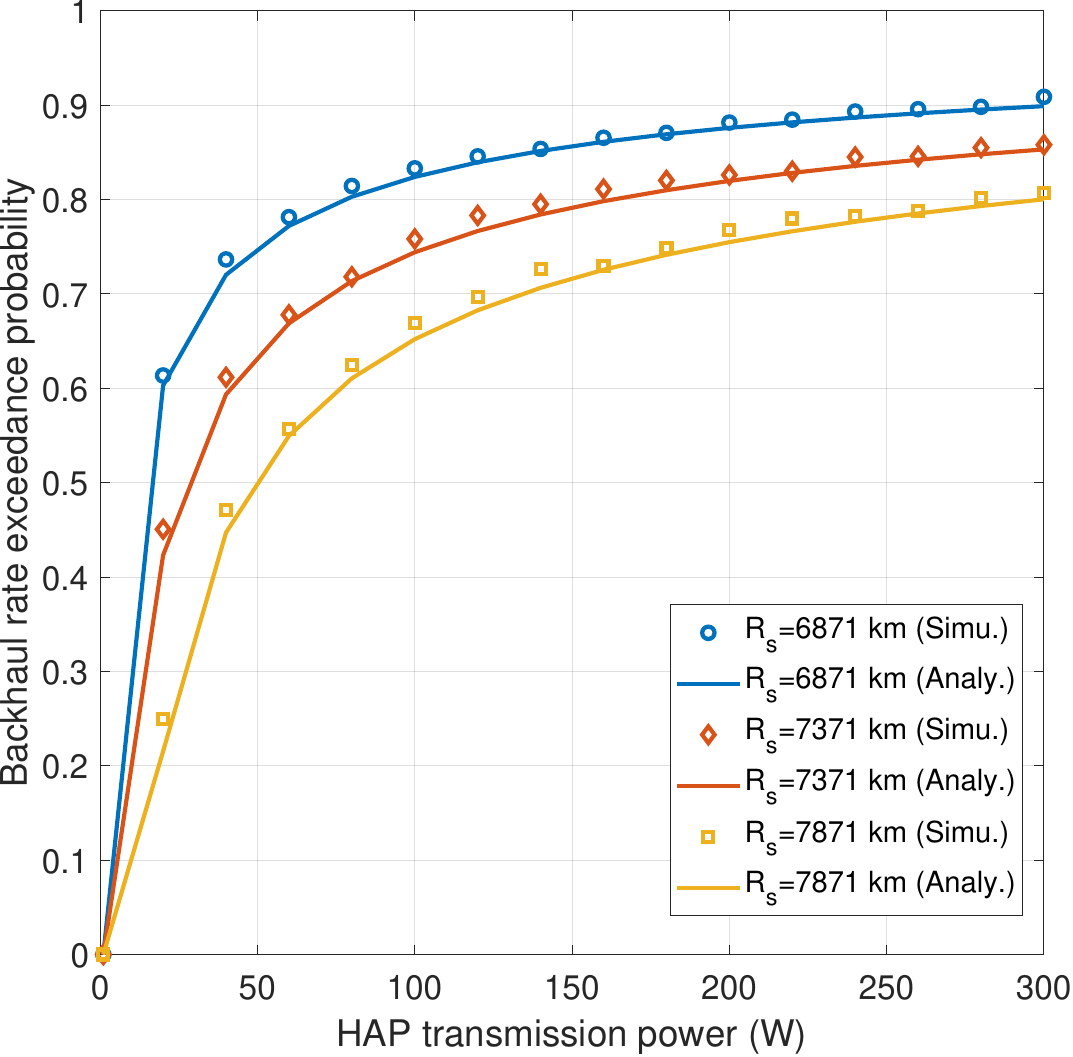}
\caption{BREP vs $P_H^t$ for different $R_s$ values.}
\label{lineRs}
\end{figure}

\begin{figure}[ht]
\centering
\includegraphics[width=0.8\linewidth]{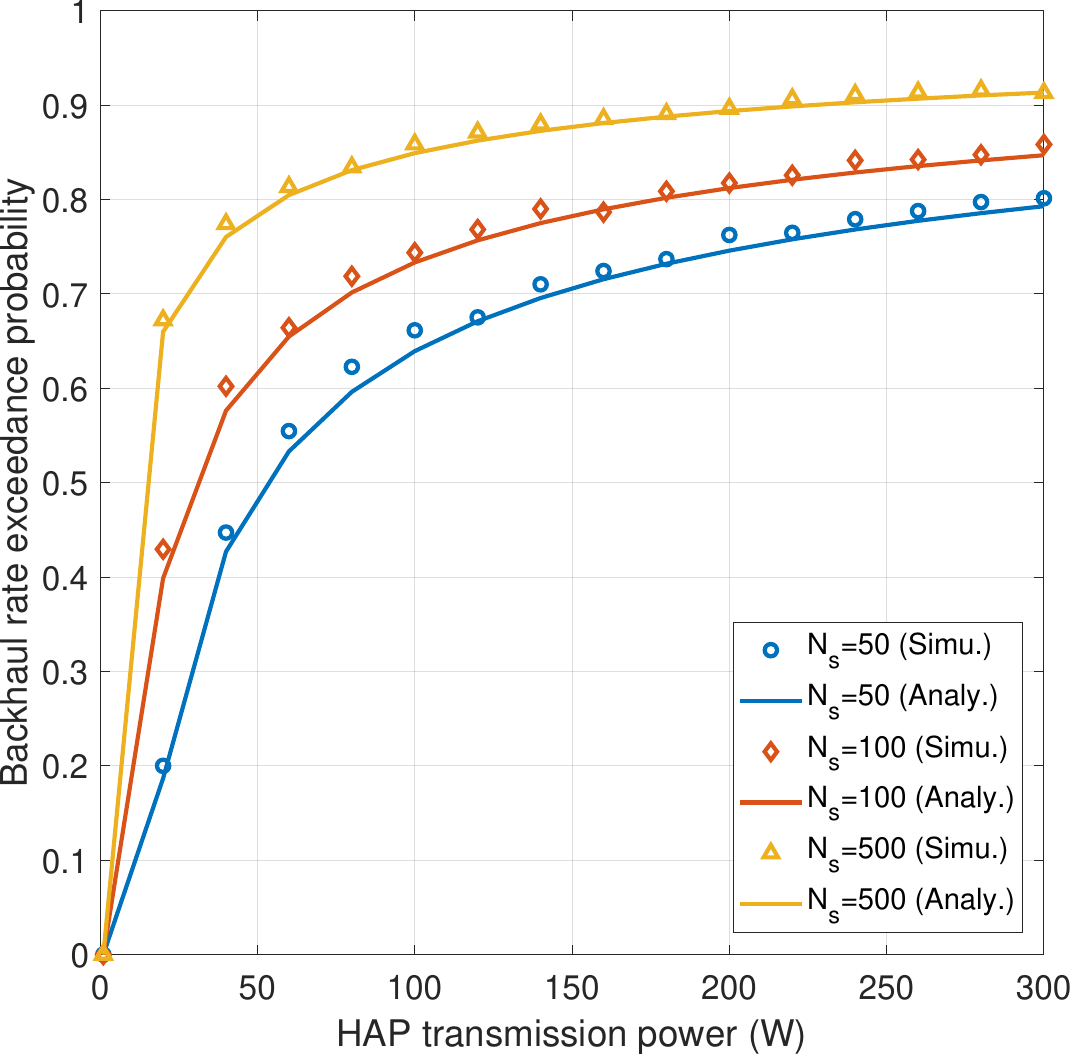}
\caption{BREP vs $P_H^t$ for different $N_s$ values.}
\label{lineNs}
\end{figure}

As seen from Figs.~\ref{lineRs} and \ref{lineNs}, there is a small gap between the analytical and simulation results, which can be explained by the logarithmic function approximation mentioned in Appendix~\ref{app:theorem3}. It is clear that the BREP curves exhibit a similar shape under different $R_s$ and $N_s$ configurations. As the HAP transmission power increases, the BREP curves first rise rapidly and then gradually converge. As the system approaches convergence, achieving higher BREP values requires a significant increase in HAP transmission power, leading to a trade-off between communication performance and energy consumption. Furthermore, reducing the satellite altitude and increasing the number of satellites can increase BERP. As a result, the current trend of building mega-constellations at low altitudes can effectively improve communication performance, whether in terms of BREP or ABDR.

\begin{figure}[ht]
\centering
\begin{minipage}{0.9\linewidth}
  \centering
  \includegraphics[width=\linewidth]{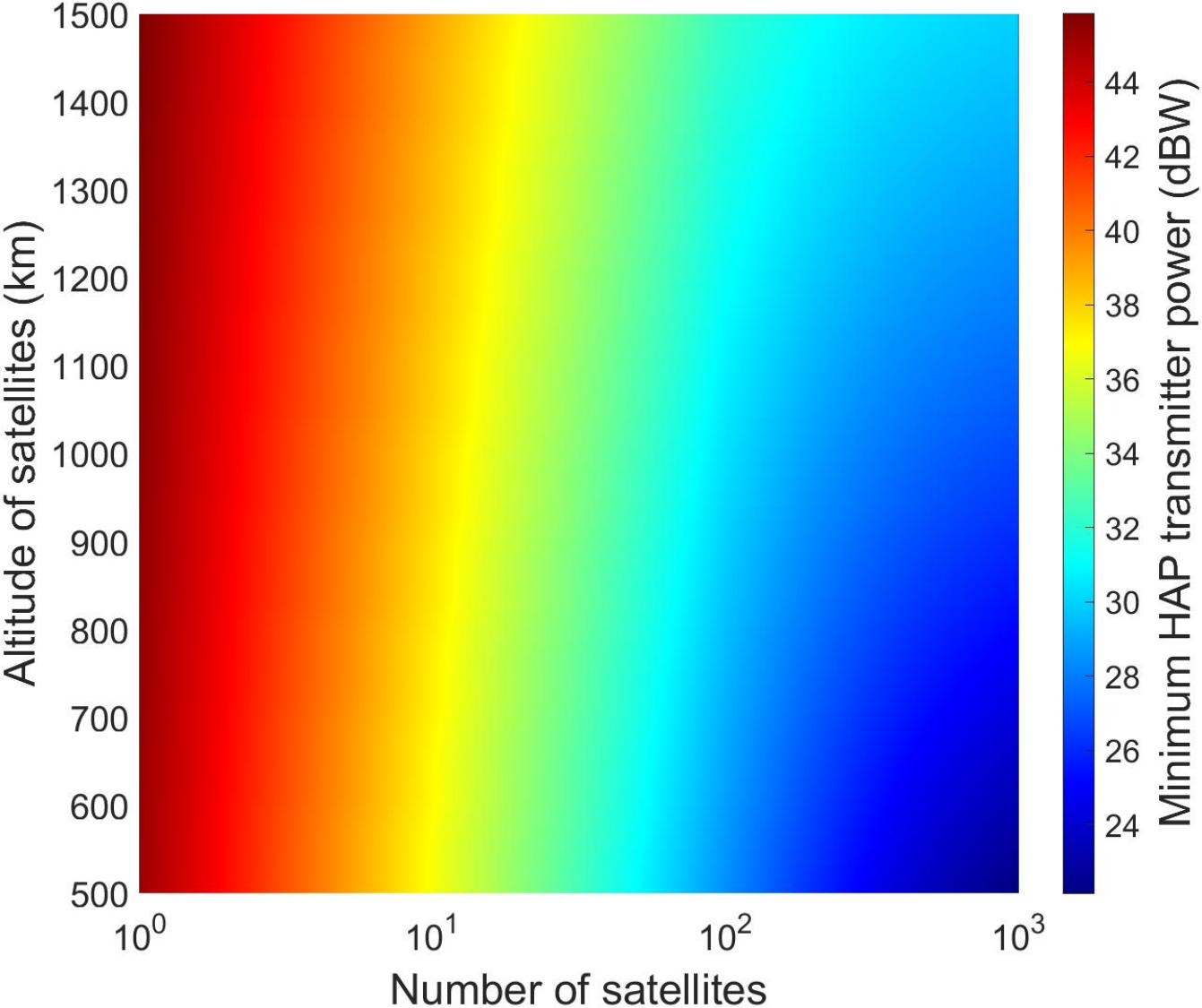}
  \captionof{subfigure}{}
  \label{heat0.9}
\end{minipage}\hfill
\begin{minipage}{0.9\linewidth}
  \centering
  \includegraphics[width=\linewidth]{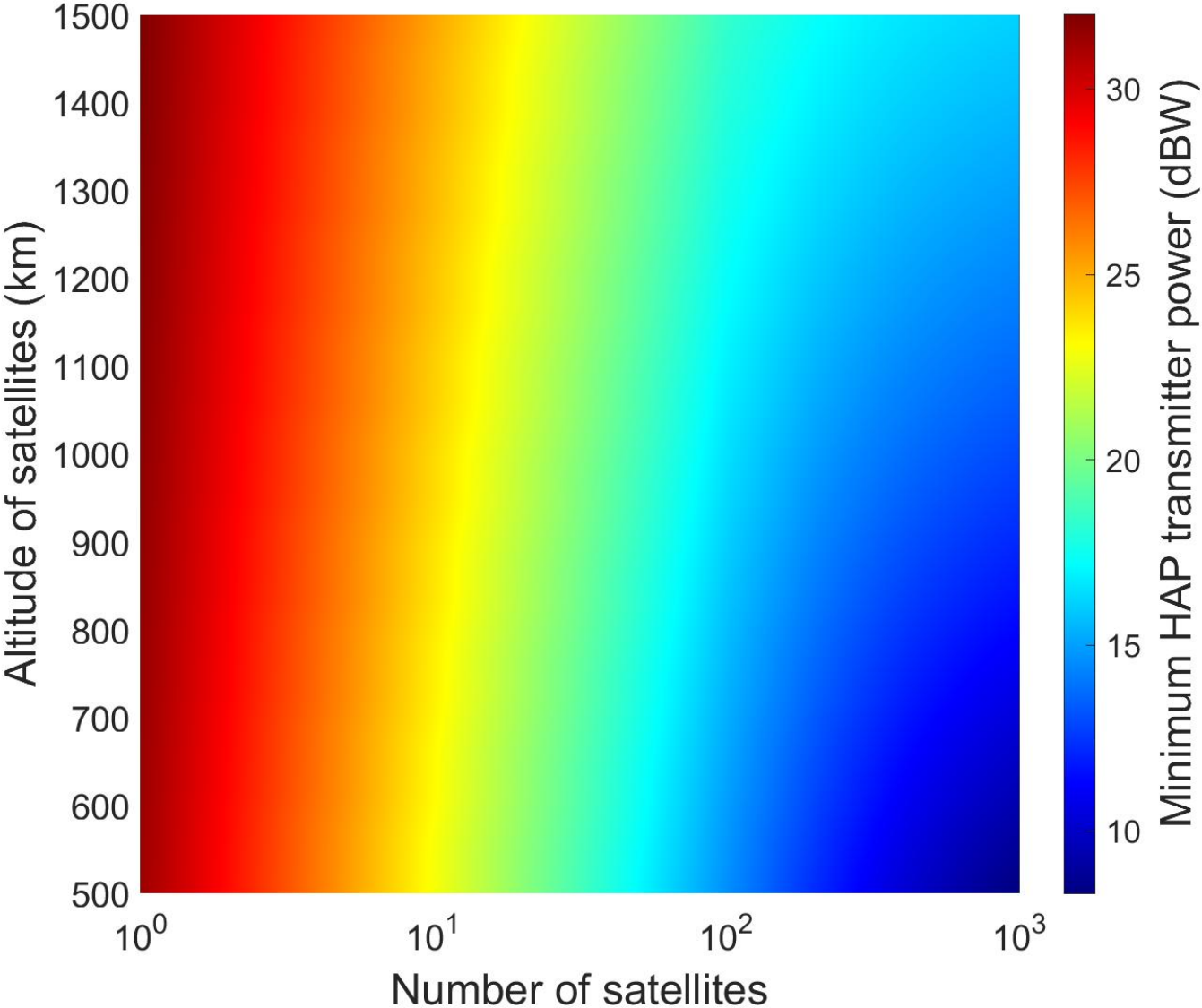}
  \captionof{subfigure}{}
  \label{heat0.5}
\end{minipage}
\caption{Minimum HAP transmission power for different constellation configurations: (a) $\mathcal{P}_{\mathrm{BREP}}=0.9$, (b) $\mathcal{P}_{\mathrm{BREP}}=0.5$.}
\label{fig:combined}
\end{figure}

\par
Fig.~\ref{fig:combined} highlights the optimal HAP transmission power to achieve $\mathcal{P}_{\mathrm{BREP}}=0.9$ and $\mathcal{P}_{\mathrm{BREP}}=0.5$ across various satellite constellations. With a constellation of satellites at an altitude of $500$~km, a BREP of $0.5$ can be achieved with a power less than $10$ dBW. However, achieving a higher exceedance probability of $0.9$ for the same configuration requires about $24$~dBW, as depicted in Fig.~\ref{heat0.9}. This indicates that while increasing the number of satellites and reducing their altitude enhances backhaul performance, achieving higher BERP requires a substantial increase in HAP transmission power.

\section{Future Research Directions}
\label{research}
This section explores promising avenues for future work, focusing on strategies for multi-HAP cooperation, hybrid communication schemes, and adaptive scheduling techniques.
\subsection{Multi-HAP Cooperation Strategies}
As future work, SAGIN's robustness can be enhanced by enabling spatial diversity and coordinated resource management across multiple HAPs. Instead of relying on a typical HAP to link both access and backhaul, as we did in this study, clusters of HAPs can be used as distributed relays. They would jointly transmit and receive signals to boost link reliability and improve spectral efficiency. By leveraging cooperative relay selection, ground users can connect through the optimal subset of HAPs, which balances traffic load and minimizes outage probability under different channel conditions. By modeling HAP positions as a PPP, one can derive analytical expressions for performance metrics in terms of HAP density, altitude, and beamwidth. Then, we can get insights into the trade-off between HAP parameters and network performance. Furthermore, the distribution of inter-HAP distances can be used to identify the optimal cluster radius for relay selection.
\subsection{Hybrid Communication Schemes}
As a future direction, we can propose an adaptive RF/FSO switching scheme for the HAP-satellite backhaul in SAGIN. The HAP will be equipped with a controller that monitors key metrics such as SNR, rain and fog attenuation, and misalignment errors, and selects the most reliable communication channel. Under high-data requirements, signals will be routed over the FSO link; when optical performance degrades, the controller will instantaneously switch to RF to preserve connectivity. This seamless handover can maximize throughput in favorable conditions and maintain continuous connectivity during adverse weather. Analytical work should be done to derive optimal switching thresholds and determine antenna designs to ensure high-performance SAGIN deployments.
\subsection{Adaptive Scheduling Methods}
As a next step, we can leverage reinforcement learning (RL) models for adaptive scheduling in a SAGIN. Rather than fixing resource allocation constraints in advance, a RL algorithm (e.g., Q-learning) can explore different scheduling decisions and receive feedback based on combined performance metrics such as throughput, delay, and link reliability. Through iterative training, the algorithm can converge on strategies that automatically balance load between access and backhaul layers and react to variations in traffic demand or channel conditions. This approach will eliminate the need for manual network parameter tuning, yielding a scheduler that continually improves its performance and maintains robust, low-latency operation under dynamic SAGIN conditions.
\section{Conclusion and Implementation Scenarios}
\label{5}
In this paper, we used tools from SSG to study the performance of a SAGIN from the relay perspective.  For this setup, we derived analytical expressions for the ABDR, AADR, and BREP. Numerical results have shown the accuracy of the analytical expressions and provided some trends of the proposed metrics by altering some network parameters, including the number and height of satellites. Some useful conclusions can be drawn from the numerical analysis, as follows. Reducing the orbital altitude of satellites and increasing their number can enhance the backhaul performance, thus creating a higher probability of securing real-time communication requirements. However, achieving higher BREP values necessitates a substantial increase in HAP transmission power, leading to a trade-off between communication performance and energy consumption. These findings underscore the importance of optimizing satellite constellation configurations and HAP transmission strategies to improve SAGIN performance.

The proposed metrics can provide valuable guidance for practical system design in SAGINs. They can be used by network planners to determine optimal relay configurations, such as selecting appropriate HAP altitudes and satellite densities based on specific reliability or data rate requirements. For instance, BREP can serve as a key design threshold to ensure that a minimum backhaul rate is achieved. This might be useful for delay-sensitive applications. Furthermore, the derived closed-form expressions can be integrated into real-time decision-making algorithms, including optimization-based frameworks or AI-driven solutions. Such methods can be used to coordinate satellite-HAP-user associations or adjust transmit powers in response to variations in user mobility or satellite configurations. As a promising future direction, reinforcement learning algorithms can be employed to learn efficient power control and relay selection mechanisms based on the insights provided by the derived performance metrics.

\appendices
\section{Proof of Lemma~\ref{lemma1}} \label{app:lemma1}
\begin{figure}[ht]
\centering
\includegraphics[width=0.7\linewidth]{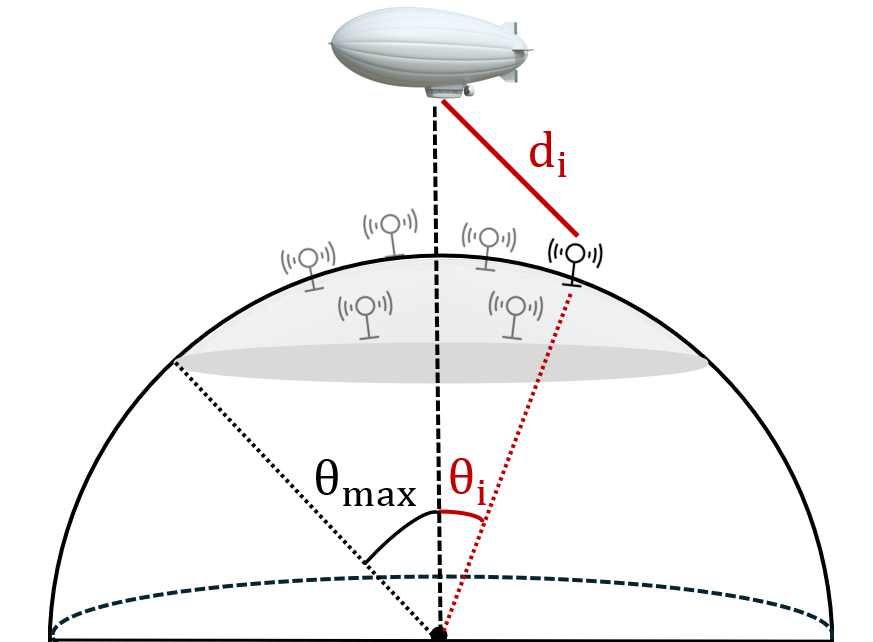}
\caption{Relation between $d_i$ and $\theta_i$.}
\label{appB}
\end{figure}

To simplify the derivation, the Euclidean distance between the $i$-th user and the typical HAP $d_i$ is presented in the form of the polar angle of the $i$-th user. An example of the relation between $d_i$ and the polar angle $\theta_i$ is shown in Fig.~\ref{appB}. The relation can be mathematically expressed according to the cosine rule: 
\begin{equation} \label{di}
d_i^2= R_\oplus^2 + R_H^2 - 2 R_\oplus R_H \cos\theta_i.
\end{equation}

\par
Then, PDF of the distribution of $d_i^2$ is derived using the Jacobian transformation:
\begin{equation}
f_{d_i^2}(d)= f_{\theta_i}(g(\theta_i))|g'(\theta_i)|,
\end{equation}
where $f_{\theta_i}(\theta)$ denotes the PDF of $\theta_i$. 
\begin{equation}
g(\theta_i)= \arccos\left(\frac{R_\oplus^2+R_H^2-d_i^2}{2R_\oplus R_H}\right),
\end{equation}
and $g'(\theta_i)$ is the derivative of $g(\theta_i)$.

\par
Referring to \cite{wang2022stochastic}, the cumulative distribution function (CDF) and PDF of $\theta_i$ are given by: 
\begin{equation}
F_{\theta_i}(\theta)= \frac{1-\cos\theta} {1-\cos\theta_{\max}},
\end{equation}
\begin{equation}
f_{\theta_i}(\theta) = \frac{\sin\theta} {1-\cos\theta_{\max}}.
\end{equation}
As a result, the PDF of $d_i^2$ will be given by:
\begin{equation}
f_{d_i^2}(d)=\frac{1}{2R_\oplus R_H (1-\cos\theta_{\max})}. 
\end{equation}
Based on the relation:
\begin{equation}
    f_{d_i}(d) = 2 d f_{d_i^2}(d^2),
\end{equation}
we can obtain the PDF of the distance distribution $f_{d_i}(d)$.

\par
Finally, we conclude the proof with the range of $d_i$. The distance from the nearest user to the HAP is always greater than the HAP's altitude $H$. Recall that the distance from the farthest user to the HAP is denoted as $d_{\max}$, and the distance bounds can be provided.

\section{Proof of Lemma~\ref{lemma2}} \label{app:lemma2}
\begin{figure}[ht]
\centering
\includegraphics[width=\linewidth]{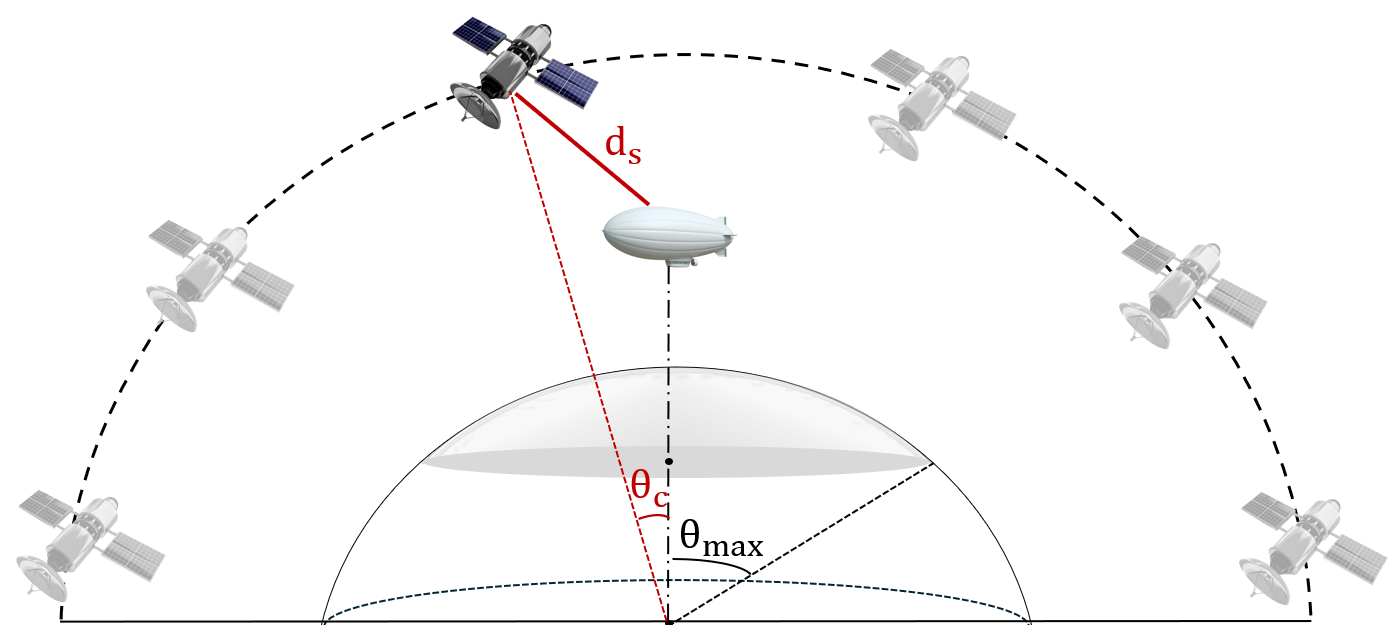}
\caption{Relation between $d_s$ and $\theta_c$.}
\label{appc}
\end{figure}

Similar to Appendix~\ref{app:lemma1}, we express $d_s$ in the form of the polar angle $\theta_c$. As shown in Fig.~\ref{appc}, the relation between them is given according to the cosine rule: 
\begin{equation}
d_s^2= R_H^2 + R_s^2- 2 R_H R_s \cos\theta_c.
\end{equation}

\par
Note that the polar angle of the nearest satellite is known as the contact angle \cite{9841569}, with the PDF given as follows:
\begin{equation}
f_{\theta_c}(\theta)= \frac{N_s}{2} \sin\theta \left( \frac{ 1+\cos\theta }{2} \right)^{N_s-1}. 
\end{equation}
The remaining steps of the derivation are similar to those of Appendix~\ref{app:lemma1}, therefore omitted here. 

\par
Finally, the shortest distance from the HAP to the satellite is the difference between their radii. To ensure that the link between the satellite and the HAP is not blocked by the Earth, the longest distance occurs when they are positioned at opposite ends of a user's horizontal plane. In this case, the distance from the HAP to the user is $\sqrt{R_H^2 - R_{\oplus}^2}$, and for the LEO satellite, it is $\sqrt{R_s^2 - R_{\oplus}^2}$. The maximum value of $d_s$ should be less than $\sqrt{R_H^2 - R_{\oplus}^2} + \sqrt{R_s^2 - R_{\oplus}^2}$.

\section{Proof of Lemma~\ref{lemma3}} \label{app:lemma3}
First of all, we expand the total received power $\mathcal{I}$ at the typical HAP into the following explicit expression: 
\begin{equation}
\mathcal{I}=\sum_{i\in \Phi_u}^{} P_u G_u G_{H}^r d_i^{-\alpha} \left(\frac{\lambda_{\mathrm{RF}}}{4\pi}\right)^2 h_{\mathrm{RF}}^2.
\end{equation}

\par
Then, the Laplace transform of $\mathcal{I}$ can be obtained as:
\begin{sequation} \label{LIs}
\begin{split}
& \mathcal{L}_{\mathcal{I}}(s) = \mathbb{E}\Big[\exp (-s\mathcal{I})|\theta, \mathcal{I} \Big] \\
& = \mathbb{E}_{\Phi_u, h_{\mathrm{RF}}^2} \Bigg[ \exp \Bigg(-s\sum_{i\in \Phi_u}^{} P_u G_u G_{H}^r d_i^{-\alpha} \left(\frac{\lambda_{\mathrm{RF}}}{4\pi}\right)^2 h_{\mathrm{RF}}^2 \Bigg) \Bigg] \\
&\stackrel{(a)}{=} \mathbb{E}_{\Phi_u} \Bigg[ \prod_{i \in \Phi_u} \mathbb{E}_{h_{\mathrm{RF}}^2} \exp \Bigg(-s P_u G_u G_{H}^r d_i^{-\alpha}\left(\frac{\lambda_{\mathrm{RF}}}{4\pi}\right)^2 h_{\mathrm{RF}}^2 \Bigg)\Bigg] \\
&\stackrel{(b)}{=} \mathbb{E}_{\Phi_u} \Bigg[ \prod_{i \in \Phi_u}  \left(1+sm_2 P_{u}G_{u}G_H^r d_i^{-\alpha}\left(\frac{\lambda_{\mathrm{RF}}}{4\pi}\right)^2 \right)^{-m_1} \Bigg].
\end{split}
\end{sequation}

\par
In (\ref{LIs}), step $(a)$ follows from the fact that the fading distribution $h_{\mathrm{RF}}^2$ of each ground user is i.i.d and independent of the point process $\Phi_u$. Step $(b)$ follows from the moment generating function (MGF) of $h_{\mathrm{RF}}^2$. 

\par
Finally, we can obtain the final result with the probability-generating functional (PGFL) of PPP, 
\begin{equation} \label{lap}
\begin{split}
\mathcal{L}_{\mathcal{I}}(s) & = \exp \Bigg(-\Lambda_u \int_0^{2\pi}\int_0^{\theta_{\max}} \Bigg(1 - \bigg( 1 + s m_2 \\
& \times (R_\oplus^2+R_H^2-2R_\oplus R_H \cos\theta )^{-\frac{\alpha}{2}} \\
& \times P_{u}G_{u}G_H^r \left(\frac{\lambda_{\mathrm{RF}}}{4\pi}\right)^2  \bigg)^{-m_1} \Bigg) 
R_\oplus^2 \sin\theta \mathrm{d}\theta \mathrm{d}\phi\Bigg).
\end{split}
\end{equation}

\section{Proof of Lemma~\ref{lemma4}} \label{app:lemma4}
First of all, recall that the random variable $\mathcal{Z}$ is defined as:
\begin{equation}
\label{Z}
\mathcal{Z} = \frac{\rho_{H,i}^r}{\sigma_H^2 + \mathcal{I}}, 
\end{equation}
where the expressions of $\rho_{H,i}^r$ and $\mathcal{I}$ are respectively given in (\ref{powerHi}) and (\ref{sump}).

\par
Thus, we derive the CCDF of $\mathcal{Z}$ according to the definition:
\begin{equation}
\begin{split}
& \overline{F}_{\mathcal{Z}}(z) = \mathbf{Pr} \Big\{\mathcal{Z}>z\Big\} = \mathbf{Pr} \Bigg\{\frac{\rho_{H,i}^r}{\sigma_H^2 + \mathcal{I}}>z \Bigg\} \\
& = \mathbf{Pr} \Bigg\{\rho_{H,i}^r>z(\sigma_H^2 + \mathcal{I})\Bigg\} \\
& = \mathbf{Pr} \Bigg\{h_{\mathrm{RF}}^2>z\left(\frac{4\pi}{\lambda_\mathrm{RF}}\right)^2\frac{1}{P_u G_u G_{H}^r}d_i^{\alpha}(\sigma_H^2 + \mathcal{I})\Bigg\} \\
& = \mathbb{E}_{\mathcal{I}} \Bigg[ 1-F_{h_{\mathrm{RF}}^2} \Bigg( \frac{zd_i^{\alpha}}{P_u G_u G_{H}^r}\left(\frac{4\pi}{\lambda_\mathrm{RF}}\right)^2(\sigma_H^2 + \mathcal{I})\Bigg) \Bigg]. 
\end{split}
\end{equation}

\par
Moreover, as a Gamma function, the CDF of $h_{\mathrm{RF}}^{2}$ can be tightly bounded by \cite{talgat2024enhancing}:
\begin{equation}
\label{upper}
    F_{h_{\mathrm{RF}}^2}(h) \leq \Bigg[ 1 - \exp \left(-\frac{(m_1!)^{-\frac{1}{m_1}}}{m_2}h\right)\Bigg]^{m_1},
\end{equation}
where $m_1 \geq 1$. Therefore, we approximate the CCDF as its tight upper bound and obtain the following result:
\begin{equation}
\begin{split}
& \overline{F}_{\mathcal{Z}}(z) \approx \mathbb{E}_{\mathcal{I}} \Bigg[1-\Bigg( 1 - \exp \Bigg( -\frac{(m_1!)^{-\frac{1}{m_1}}}{m_2} \\ 
& \times \frac{zd_i^{\alpha}}{P_u G_u G_{H}^r} \left(\frac{4\pi}{\lambda_\mathrm{RF}}\right)^2(\sigma_H^2 + \mathcal{I})\Bigg) \Bigg)^{m_1} \Bigg].
\end{split}
\end{equation}

\par
For simplicity of expression, we defined $\beta$ and $\mu(d_i)$ as: 
\begin{align}
\beta & =\left((m_1)!\right)^{-\frac{1}{m_1}}, \\
\mu\left(d_i\right) & =\frac{zd_i^{\alpha}}{m_2P_u G_u G_{H}^r}\left(\frac{4\pi}{\lambda_\mathrm{RF}}\right)^2, 
\end{align}
and $\overline{F}_{\mathcal{Z}}(z)$ can be expressed as
\begin{sequation}
\overline{F}_{\mathcal{Z}}(z) \approx  \mathbb{E}_{\mathcal{I}} \left[1-\left( 1 - \exp \left( - \beta \mu\left(d_i\right) (\sigma_H^2 + \mathcal{I})\right) \right)^{m_1} \right].
\end{sequation}

Using Newton's generalized binomial theorem and taking the expectation with respect to $\mathcal{I}$, the proof is finished:
\begin{equation}
\begin{split}
\overline{F}_{\mathcal{Z}}(z) & \approx  \mathbb{E}_{\mathcal{I}} \left[1-\left( 1 - \exp \left( - \beta \mu\left(d_i\right) (\sigma_H^2 + \mathcal{I})\right) \right)^{m_1} \right] \\
& = \mathbb{E}_{\mathcal{I}} \Bigg[1- \sum_{k=0}^{m_1} \binom{m_1}{k} (-1)^{k} \\
& \times \exp\bigg(-k \beta \mu(d_i) (\sigma_H^2 + \mathcal{I}) \bigg)\Bigg] \\
& = \sum_{k=1}^{m_1} \binom{m_1}{k} (-1)^{k+1} \\
& \times \exp(-k\beta\mu(d_i)\sigma_H^2) \mathcal{L}_{\mathcal{I}}\bigg(k \beta \mu(d_i)\bigg).
\end{split}
\end{equation}

\section{Proof of Theorem~\ref{theorem1}} \label{app:theorem1}
Note that the received power from each user does not differ significantly. Given that $\gamma_{H,i} \ll 1$, the approximation
\begin{equation}
    \ln(1+\gamma_{H,i}) \approx \gamma_{H,i}
\end{equation}
holds. Then, we start the derivation of AADR:
\begin{equation}
\begin{split}
& \mathcal{R}_{\mathrm{RF}} =  \mathbb{E} \Bigg[\sum_{i\in \Phi_u} B_{\mathrm{RF}}\log_2\left(1+\gamma_{H,i}\right)\Bigg] \\
& \approx \frac{B_{\mathrm{RF}}}{\ln2} \,
\mathbb{E} \Bigg[\sum_{i\in \Phi_u} \gamma_{H,i} \Bigg] = \frac{B_{\mathrm{RF}}}{\ln2} \mathbb{E} \Bigg[\sum_{i\in \Phi_u} \frac{\rho_{H,i}^r}{\sigma_H^2 + \mathcal{I}_i} \Bigg].
\end{split}
\end{equation}

\par
Since $\gamma_{H,i}$ is dependent on both user $i$ and the point process $\Phi_u\setminus \{i\}$, the standard Campbell's theorem cannot be applied. The dependence of $\gamma_{H,i}$ on $\Phi_u\setminus \{i\}^{}$ is due to interference $\mathcal{I}_i$. Therefore, we use the closely related Campbell-Mecke theorem, which for a function $f(X_i, \Phi\setminus\{X_i\})$ gives: 
\begin{equation}
\mathbb{E} \Bigg[\sum_{X_i\in \Phi} f(X_i, \Phi\setminus\{X_i\}) \Bigg]= \Lambda \int_{\mathbb{R}^{d}} \mathbb{E}\Big[f(x_i, \Phi)\Big] \mathrm{d} x_i.
\end{equation}
Therefore, we have
\begin{sequation}
\begin{split}
\mathcal{R}_{\mathrm{RF}} & = \frac{\Lambda_u B_{\mathrm{RF}}}{\ln2} \int_{0}^{2\pi} \int_{0}^{\theta_{\max}} \mathbb{E} \Bigg[\frac{\rho_{H,i}^r}{\sigma_H^2 + \mathcal{I}} \Bigg] R_\oplus^2 \sin\theta \mathrm{d}\theta \mathrm{d}\phi \\
& = \frac{2\pi \Lambda_u R_\oplus^{2} B_{\mathrm{RF}}}{\ln2}\int_{0}^{\theta_{\max}}  
\Bigg(\int_{0}^{1} \overline{F}_{\mathcal{Z}}(z) \, dz \Bigg) \sin\theta \mathrm{d}\theta, 
\end{split}
\end{sequation}
where the second step stands by using 
\begin{equation}
\mathbb{E} [\mathcal{Z}] = \int_{0}^{1} \Pr \Big\{\mathcal{Z} > z\Big\} \mathrm{d}z. 
\end{equation}

\section{Proof of Lemma~\ref{lemma5}} \label{app:lemma5}
Given the distribution of the random deviation angle $\theta_d$, the PDF of $h_{\mathrm{FSO}}$ can be obtained as:
\begin{equation}
\begin{split}
& f_{h_{\mathrm{FSO}}} \left(h\right) = \int_0^{\infty} f_{h_{\mathrm{FSO}}|\theta_d}\left(h\right) f_{\theta_d}\left(\theta\right) \mathrm{d} \theta \\
&= \frac{\eta_s^2}{A_0^{\eta_s^2} \sigma_0^2} h^{\eta_s^2-1} 
\int_0^{\infty} \theta \cos\theta \exp\left(-\frac{\theta^2}{2\sigma_0^2}\right) \mathrm{d} \theta \\
&\stackrel{(a)}{=} \frac{\eta_s^2}{A_0^{\eta_s^2}} h^{\eta_s^2-1} 
\exp\left(-\frac{\sigma_0^2}{2}\right) {}_1F_1\left( \frac{-1}{2}, \frac{1}{2}, \frac{\sigma_0^2}{2} \right),    
\end{split}
\end{equation}
where $0 \leq h \leq A_0$, and ${}_1F_1( \cdot, \cdot, \cdot )$ is the Hypergeometric function. Step $(a)$ follows from: 
\begin{equation}
\begin{split}
& \int_0^{\infty} x^{\mu-1} e^{-\beta x^2} \cos(ax) \mathrm{d} x \\
& = 
\frac{1}{2} \beta^{-\frac{\mu}{2}} \Gamma\left(\frac{\mu}{2}\right) e^{\frac{-a^2} {4\beta}} 
{}_1F_1\left( \frac{1-\mu}{2}, \frac{1}{2}, \frac{a^2}{4\beta} \right),
\end{split}
\end{equation}
where $a>0$, $Re(\mu)>0$, $Re(\beta)>0$.

Then, the CDF of $h_{\mathrm{FSO}}$ is as follows:
\begin{equation}
\label{CDFFSO}
F_{h_{\mathrm{FSO}}}\left(h\right) = \frac{1}{A_0^{\eta_s^2}} h^{\eta_s^2} 
\exp\left(-\frac{\sigma_0^2}{2} \right) {}_1F_1\left( \frac{-1}{2}, \frac{1}{2}, \frac{\sigma_0^2}{2} \right).
\end{equation}

\section{Proof of Theorem~\ref{theorem2}} \label{app:theorem2}
We obtain the ABDR by taking the expectation with respect to $d_s$ and $h_{\mathrm{FSO}}$:
\begin{sequation}
\begin{split}
& \mathcal{R}_{\mathrm{FSO}} = \frac{B_{\mathrm{FSO}}}{\ln2}\mathbb{E}_{d_s,h_{\mathrm{FSO}}} \Bigg[\ln \left(1+ \frac{\upsilon^2 P_H^t G_H^t G_s \lambda_{\mathrm{FSO}}^2 h_{\mathrm{FSO}}^2}{(4\pi)^2 \sigma_s^2 d_s^2} \right) \Bigg] \\
& = \frac{B_{\mathrm{FSO}}}{\ln2}\mathbb{E}_{d_s} \Bigg[\int_{0}^{A_0}\ln \left(1+ \frac{\upsilon^2 P_H^t G_H^t G_s \lambda_{\mathrm{FSO}}^2 h^2}{(4\pi)^2 \sigma_s^2 d_s^2} \right) f_{h_\mathrm{FSO}}(h)\Bigg],
\end{split}
\end{sequation}
where the second step stands because of the independence of $d_s$ and $h_{\mathrm{FSO}}$. Finally, taking the expectation with respect to $d_s$ completes the proof.

\section{Proof of Theorem~\ref{theorem3}} \label{app:theorem3}
We start the derivation of BREP from its definition, and the derivation details are presented in 
(\ref{long}) at the top of the next page.
\begin{table*}
\begin{equation}\label{long}
\begin{split}
& \mathcal{P}_{\mathrm{BREP}} = \mathbf{Pr} \Big\{B_{\mathrm{FSO}}\log_2\left(1+\gamma_s\right)>B_{\mathrm{RF}}\sum_{i\in \Phi_u} \log_2\left(1+\gamma_{H,i}\right)\Big\}  \\ 
&\stackrel{(a)}{\approx} \mathbf{Pr} \Bigg\{\gamma_s>\exp \left(\frac{B_{\mathrm{RF}}}{B_{\mathrm{FSO}}}\sum_{i\in \Phi_u} \gamma_{H,i}\right) -1 \Bigg\}  \\ 
&\stackrel{(b)}{=} \mathbf{Pr} \Bigg\{\frac{\upsilon^2 P_H^t G_H^t G_s \lambda_{\mathrm{FSO}}^2h_{\mathrm{FSO}}^2}{(4\pi)^2 \sigma_s^2 d_s^2} > \exp \left(\frac{B_{\mathrm{RF}}}{B_{\mathrm{FSO}}} \sum_{i\in \Phi_u} \gamma_{H,i}\right) -1 \Bigg\}    \\ 
& \stackrel{(c)}{=} \mathbf{Pr}\Bigg\{h_{\mathrm{FSO}} > \varepsilon d_s\sqrt{\exp \Big({\frac{B_\mathrm{RF}}{B_\mathrm{FSO}} \sum_{i\in \Phi_u} \gamma_{H,i}}\Big)-1}\Bigg\}   \\ 
& =\mathbb{E}_{d_s,\Phi_u}\Bigg[1 - F_{h_\mathrm{FSO}}\Bigg(\varepsilon d_s\sqrt{\exp \Big({\frac{B_\mathrm{RF}}{B_\mathrm{FSO}} \sum_{i\in \Phi_u} \gamma_{H,i}}\Big)-1}\Bigg)\Bigg]   \\ 
& \stackrel{(d)}{=}  1-\frac{1}{A_0^{\eta_s^2}}\exp\left(-\frac{\sigma_0^2}{2}\right) \, {}_1F_1\left(-\frac{1}{2}, \frac{1}{2}, \frac{\sigma_0^2}{2}\right) \, \varepsilon^{\eta_s^2} \mathbb{E}_{d_s}[d_s^{\eta_s^2}]  \mathbb{E}_{\Phi_u} \Bigg[ \Bigg(\exp \Big({\frac{B_\mathrm{RF}}{B_\mathrm{FSO}} \sum_{i\in \Phi_u} \gamma_{H,i}}\Big)-1\Bigg)^{\frac{\eta_s^2}{2}}\Bigg]   \\
&=  1-\frac{1}{A_0^{\eta_s^2}}\exp\left(-\frac{\sigma_0^2}{2}\right) \, {}_1F_1\left(-\frac{1}{2}, \frac{1}{2}, \frac{\sigma_0^2}{2}\right) \, \varepsilon^{\eta_s^2} \mathbb{E}_{d_s}[d_s^{\eta_s^2}]  \mathbb{E}_{\Phi_u} \Bigg[ \exp \Big({\frac{\eta_s^2}{2}\frac{B_\mathrm{RF}}{B_\mathrm{FSO}} \sum_{i\in \Phi_u} \gamma_{H,i}}\Big)\Bigg(1-\exp \Big({-\frac{B_\mathrm{RF}}{B_\mathrm{FSO}} \sum_{i\in \Phi_u} \gamma_{H,i}}\Big)\Bigg)^{\frac{\eta_s^2}{2}}\Bigg]   \\
& \stackrel{(e)}{=}  1-\frac{1}{A_0^{\eta_s^2}}\exp\left(-\frac{\sigma_0^2}{2}\right) \, {}_1F_1\left(-\frac{1}{2}, \frac{1}{2}, \frac{\sigma_0^2}{2}\right) \, \varepsilon^{\eta_s^2} \mathbb{E}_{d_s}[d_s^{\eta_s^2}] \\
& \times \mathbb{E}_{\Phi_u} \Bigg[ \exp \Big({\frac{\eta_s^2}{2}\frac{B_\mathrm{RF}}{B_\mathrm{FSO}}\sum_{i\in \Phi_u} \gamma_{H,i}}\Big) \sum_{n=0}^{\infty} (-1)^n \frac{\Gamma(\frac{\eta_s^2}{2}+1)}{n! \, \Gamma(\frac{\eta_s^2}{2}-n+1)} \exp \Big({-n\frac{B_\mathrm{RF}}{B_\mathrm{FSO}} \sum_{i\in \Phi_u} \gamma_{H,i}}\Big) \Bigg]   \\
& \stackrel{(f)}{=} 1 - \frac{1}{A_0^{\eta_s^2}} \exp\left(-\frac{\sigma_0^2}{2}\right) 
{}_1F_1\left(-\frac{1}{2}, \frac{1}{2}, \frac{\sigma_0^2}{2}\right) 
\varepsilon^{\eta_s^2} \mathbb{E}_{d_s}[d_s^{\eta_s^2}]  \mathbb{E}_{\Phi_u} \Bigg[\sum_{n=0}^{\infty} (-1)^n 
\frac{\Gamma\left(\frac{\eta_s^2}{2}+1\right)}{n! \, \Gamma\left(\frac{\eta_s^2}{2}-n+1\right)}
\exp \left(\frac{B_\mathrm{RF}}{B_\mathrm{FSO}}\left(\frac{\eta_s^2}{2}-n\right) \sum_{i\in \Phi_u} \gamma_{H,i} \right) \Bigg].     
\end{split}
\end{equation}
\rule{\linewidth}{0.4pt}
\end{table*}
In (\ref{long}), step $(a)$ comes from the approximation $\ln(1+\gamma_{H,i}) \approx \gamma_{H,i}$ and $(b)$ follows from the expansion of $\gamma_s$. In step $(c)$, the expression $\varepsilon$ is given by:
\begin{equation}
\varepsilon=\frac{4 \pi \sigma_s}{\lambda_\mathrm{FSO} v \sqrt{P_H^t G_H^t G_s}},
\end{equation}

\par
Next, step $(d)$ follows from using (\ref{CDFFSO}) and the independence between $d_s$ and the process $\Phi_u$. Finally, we obtain $(e)$ by using Newton's generalized binomial theorem for any real number $r$, given by:
\begin{equation}
(1 + z)^r = \sum_{n=0}^{\infty} \frac{\Gamma(r+1)}{n! \, \Gamma(r - n + 1)} z^n,
\label{newt}
\end{equation}
which converges for $|z|<1$.

In case of having a non-negative integer r, we can apply directly the classical binomial identity given by:
\begin{equation}
(1 + z)^r= \sum_{n=0}^{r} \binom{r}{n} z^n.
\end{equation}

Then, we use the linearity of the expectation and get
\begin{sequation}
\begin{split}
&\mathbb{E}_{\Phi_u} \Bigg[\sum_{n=0}^{\infty} (-1)^n 
\frac{\Gamma\left(\frac{\eta_s^2}{2}+1\right)}{n! \, \Gamma\left(\frac{\eta_s^2}{2}-n+1\right)}e^{\frac{B_\mathrm{RF}}{B_\mathrm{FSO}}\left(\frac{\eta_s^2}{2}-n\right) \sum_{i\in \Phi_u} \gamma_{H,i}} \Bigg]   \\
& = \sum_{n=0}^{\infty} (-1)^n 
\frac{\Gamma\left(\frac{\eta_s^2}{2}+1\right)}{n! \, \Gamma\left(\frac{\eta_s^2}{2}-n+1\right)}\mathbb{E}_{\Phi_u} \Bigg[ e^{\frac{B_\mathrm{RF}}{B_\mathrm{FSO}}\left(\frac{\eta_s^2}{2}-n\right)\sum_{i\in \Phi_u} \gamma_{H,i}}\Bigg]   \\
& = \sum_{n=0}^{\infty} (-1)^n 
\frac{\Gamma\left(\frac{\eta_s^2}{2}+1\right)}{n! \, \Gamma\left(\frac{\eta_s^2}{2}-n+1\right)}
\mathbb{E}_{\Phi_u} \left[ \prod_{i \in \Phi_u} e^{ \frac{B_\mathrm{RF}}{B_\mathrm{FSO}}\left(\frac{\eta_s^2}{2}-n\right) \gamma_{H,i}} \right].
\end{split}
\end{sequation}

In step (f), the convergence of the infinite series can be explained as follows:
\begin{equation}
\begin{split}
&\sum_{n=0}^{\infty} (-1)^n 
\frac{\Gamma\left(\frac{\eta_s^2}{2}+1\right)}{n! \, \Gamma\left(\frac{\eta_s^2}{2}-n+1\right)}
\exp \left(\frac{B_\mathrm{RF}}{B_\mathrm{FSO}}\left(\frac{\eta_s^2}{2}-n\right) \sum_{i\in \Phi_u} \gamma_{H,i} \right)\\
&= \exp \Big({\frac{\eta_s^2}{2}\frac{B_\mathrm{RF}}{B_\mathrm{FSO}} \sum_{i\in \Phi_u} \gamma_{H,i}}\Big)\Bigg(1-\exp ({-\frac{B_\mathrm{RF}}{B_\mathrm{FSO}} \sum_{i\in \Phi_u} \gamma_{H,i}}\Big)\Bigg)^{\frac{\eta_s^2}{2}}.
\end{split}
\end{equation}

Based on the convergence condition of Newton's generalized binomial theorem applied in (\ref{newt}), we have:
\begin{equation}
|-e^{{-\frac{B_\mathrm{RF}}{B_\mathrm{FSO}} \sum_{i\in \Phi_u} \gamma_{H,i}}}|<1. 
\end{equation}
Therefore, the infinite series under investigation is convergent.

Finally, the proof is concluded with the closely related Campbell-Mecke theorem:
\begin{sequation}
\begin{split}
& \sum_{n=0}^{\infty} (-1)^n 
\frac{\Gamma\left(\frac{\eta_s^2}{2}+1\right)}{n! \, \Gamma\left(\frac{\eta_s^2}{2}-n+1\right)}
\mathbb{E}_{\Phi_u} \left[ \prod_{i \in \Phi_u} e^{ \frac{B_\mathrm{RF}}{B_\mathrm{FSO}}\left(\frac{\eta_s^2}{2}-n\right) \gamma_{H,i}} \right] \\
& = \sum_{n=0}^{\infty} (-1)^n \frac{\Gamma\left(\frac{\eta_s^2}{2}+1\right)}{n! \, \Gamma\left(\frac{\eta_s^2}{2}-n+1\right)}\exp \Big(-2\pi \Lambda_u R_{\oplus}^2   \\
& \times \int_{0}^{\theta_{\max}} \Big(1-\mathbb{E}\Big[e^{\frac{B_\mathrm{RF}}{B_\mathrm{FSO}}(\frac{\eta_s^2}{2}-n)\mathcal{Z}}\Big]\Big)\sin\theta d\theta\Big).
\end{split}
\end{sequation}

\bibliographystyle{IEEEtran}
\bibliography{sample}

\vfill\pagebreak

\end{document}